\documentclass[%
 preprint,
 amsmath,amssymb,
 aip,
12pt]{revtex4-2}

\usepackage{graphicx}
\usepackage{dcolumn}
\usepackage{bm}
\usepackage{comment}
\usepackage{svg}
\usepackage{appendix}
\usepackage{amsmath}
\usepackage{xparse}
\usepackage{physics}
\usepackage{times}
\usepackage{graphicx}
\usepackage{array}
\usepackage[normalem]{ulem}

\usepackage{cmbright}
\DeclareFontShape{OT1}{cmss}{m}{it}{<->ssub*cmss/m/sl}{}

\begin{document}


\title{Numerical study of airborne particle dynamics in vortices subject to electric field}

\author{Pramodt Srinivasula}
    \email{pramodt@iitb.ac.in}
\author{Rochish M Thaokar}%

\affiliation{%
 Indian Institute of Technology Bombay, Mumbai\\
Maharashtra, 400076, India}

\date{\today}

\begin{abstract}
Capture, selective collection and flight manipulation of airborne particulate are three important functional requirements in various actively growing aerosol technology applications. Aerodynamic drag, particle inertia and dielectrophoretic (DEP) force due to externally applied electrostatic forces influence the behavior of micron sized particles significantly, in such situations. In this work, we numerically study how a combination of these forces uniquely influences the behavior of uncharged or mildly charged airborne particles, distinct from that with their individual influences. Uncharged particle movements in a numerically fabricated well structured steady vortical flow between two curved electrode surfaces are analyzed. Vortical air circulation towards and away from the electrode tip enhances and deteriorates electrostatic particulate capture on the electrodes, termed as co and counter directions with respect to the electrostatic force on particles respectively. Although particle behavior is monotonous in co-direction vortical flows, particles in counter vortices subject to an electric field reveals a rich variety of unique behaviors attributing to the interplay of drag, inertia and DEP forces based on their relative magnitudes.
Distortion of the vortex structure due to convexity of electrode surfaces results in an \textit{inverse inertial limit cycle trajectory trapping} of particles; with the airborne particles spatially segregated, trapping larger particles further inside the vortex than the smaller particles. Value of a dimensionless number $\xi_v$, the ratio of DEP force and particle inertia, represents the combination of the operating flow and electric field strengths. Inertial cut-off of particle capture in this configuration abruptly shift to DEP capture at a critical value of $\xi_v\approx 0.2$, as its value increases. Selective deposition of particles within a closed range of size and density, emerges due to the interplay of vortex trapping, inertial expulsion and electrostatic capture.
\end{abstract}

\maketitle
\section{\label{sec:intro}Introduction}
Suspended particulate dynamics in air circulations past surfaces has been studied widely in both naturally occurring and human made systems\cite{twomey1977atmospheric,acheson1991elementary}. Airborne particles behavior in vortex flows and electrostatic fields has also been studied over several decades for airborne particle separation\cite{davies1953separation,rhodes2008introduction,pich2017gas,friedlander1977smoke}. Combined influence of strong ambient gas flow circulations and electric fields on micron sized airborne particles arises widely in such situations. Such as, aerosol droplets in turbulent flow circulations of thundering clouds\cite{colgate1967enhanced,gabyshev2020condensational} and particles in gas flow through electrostatic precipitators (ESP)\cite{yamamoto1981electrohydrodynamics,lu2017analysis}. ESPs are popular for industrial gas filtration due to low flow restriction to the gas flow, as compared to fiber or membrane filters\cite{jaworek2007modern,parker1997electrostatic,mizuno2000electrostatic}. However such ESPs ionize air to impart high charge onto particulate for improving their capture, which has serious health and environmental concerns\cite{boelter1997ozone,chen2002ozone}. Hence, high efficiency of capture of fine particulate without ionizing air is an active technological pursuit in the recent years, especially for air cleaning in indoor environments\cite{tepper2007electrospray}.

Modern applications such as selective aerosol sampling and active aerosol delivery are on the rise in the recent years. Fast, reliable and controllable active selective collection of airborne biological matter without injuring their cells, was recently recognized as an essential yet challenging research problem for bioaerosol investigation in medical, biological and environmental sciences\cite{mainelis2020bioaerosol}. In this fascinating pursuit of developing such aerosol collection techniques suitable for a wide range of airborne matter and ambient conditions, researchers are shifting their focus\cite{mainelis2020bioaerosol} from conventional inertial impactors \cite{andersen1958new,nevalainen1992performance} and interception based passive fiber filters to exploring more possible techniques \cite{mainelis2020bioaerosol}. With the advancements in synthetic aerosol drug development, pulmonary drug delivery techniques are emerging as a promising medical technology over the past decade\cite{dolovich2011aerosol,pleasants2018aerosol}. Few current technological bottlenecks and opportunities for further scientific research in these areas of airborne particulate capture, selective sampling and flight manipulation are identified below.
\section{\label{sec:motivation}Motivation: Aerodynamic and electrostatic influences on particulate} 
Particle size and inertia primarily determines the relative dominance of various influences on it. Thermal diffusion of airborne particles is important for particles of sizes below 0.1micron. Air flow drag significantly influences typical fine ambient airborne particles of near micron sizes\cite{owen1992airborne,hinds1999aerosol,pich2017gas,friedlander1977smoke}. Newton's drag force formulation in terms of a drag coefficient $C_D$ accounts for inertial resistance of high Reynolds number ($Re_p$) flow of ambient fluid relative to the particle. At low Reynolds number Stokes's formulation accounts for viscous resistance from gas. Drag force on particles in these two formulations scale with $v^2 d_p^2$ and $v d_p$ respectively; Where $d_p$ is the particle aerodynamic diameter, $v$ is magnitude of relative velocity between particle and gas. Although intermediate $Re_p$ cases and influence of particle shape are captured by more sophisticated models, expressed typically in terms of $C_D$ and $Re$, Stokes formulation holds satisfactorily for a wide range of laminar gas flows\cite{morsi1972investigation,hinds1999aerosol}. Particle inertia and electrostatic forces are significant for particle size, typically above 10micron.

Particles dispersed in a gas under a constant electric field are acted upon primarily by two electrostatic forces, dielectrophoretic (DEP) force and Coulombic force. Relative polarization of particle in gas in presence of a non uniform electric field $\textbf{E}$ across the particle results in DEP force expressed as, $\textbf{F}_{dep}= \frac{\pi}{4}d_p^3 \epsilon_0 \epsilon_a \kappa_{cm}\nabla |\textbf{E}|^2$. Where $\kappa_{cm}= \frac{\epsilon_p-\epsilon_a}{\epsilon_p+2 \epsilon_a}$ is the Clausius-Mossotti factor representing the relative polarizability of particles in the bulk fluid. $\epsilon_0, \epsilon_a, \epsilon_p$ are the permittivity of free space, dielectric constants of bulk fluid and particles respectively. 
Coulombic force on a particle with a net charge $q_p$ in the electric field is, $\textbf{F}_C = q_p \textbf{E}$. 
Air flow velocity, electric field and the electric field gradient in typical wire-plate configurations ESPs are typically of the order of $1m/s$,$10 kV/cm$ and $200 kV/cm^2$ respectively. \cite{yamamoto1981electrohydrodynamics}. Figure \ref{fig:DEP_dragForces} in the appendix \ref{append:crescentModel}, indicates how dominance of drag force on fine micron sized particles shifts to that of electrostatic forces for larger sizes of ambient uncharged or mildly charged particles, at such flow and electric field magnitudes. 
Curved electrode surfaces (such as wire-plate) are often used, to increase the non-uniformity of electric field to enhance DEP force. Similarly, to improve the Coulombic force on the particles for a better capture, thousands of electrons or ions are imparted on them using a plasma discharge arc in ESPs\cite{yamamoto1981electrohydrodynamics,mainelis2002effect}.
 
Ionic wind or corona wind of ionized air in this process\cite{yamamoto1981electrohydrodynamics} results in a secondary flow vortical air circulations near the electrode surfaces\cite{yamamoto1981electrohydrodynamics,liang1994characteristics, podlinski2006electrohydrodynamic, chun2007numerical} and controls turbulence\cite{atten1987electrohydrodynamic,soldati1998turbulence}. Mechanically generated secondary air flow vortical circulations due to the shape of electrode were reported to improve the capture of fine micron sized particles\cite{zhu2019numerical,lu2017analysis}. However, deterioration of capture performance of fine micron particles due to gas EHD driven secondary circulations were reported\cite{liang1994characteristics,podlinski2006electrohydrodynamic,zhu2019numerical,gao2020effect,lu2017analysis}. On the other hand, Ozone and other harmful oxidized gases are released during air ionization, which are of greater concern to human health and environment\cite{britigan2006quantification,boelter1997ozone,yanallah2009experimental,chen2002ozone} and hence shall be avoided. 
To eliminate air ionization while imparting charge on to the particles, wet ESP was introduced with electrospray of charged micro droplets of water\cite{tepper2007electrospray,jaworek2007modern}. This induce a jet of $20\mu m$ to $30 \mu m$ droplets at a velocity of 16m/s, that is about 20 times higher than inlet gas flow velocity\cite{kim2010electrospray}. Hence air circulations created by the drag force the spray droplets was recognized to be significant\cite{arumugham2013two}. 

Bioaerosol sampling is another application where finite inertia of particulate and electrostatic forces are leveraged separately to collect airborne matter\cite{mainelis2020bioaerosol,haig2016bioaerosol,jing2013microfluidic,collins2017selective,khojah2017size,mainelis2002induction}. 
Selective separation and collection of the airborne particulate based on their size (typically $5-10\mu m$ aerodynamic diameter) and inertia for further transfer is required in a variety of applications of bioaerosol detection and sampling\cite{kim2007micromachined,kang2014real,kauppinen1989static}. Microfluidic device designs based on either particle inertia\cite{hong2015continuous,schaap2012transport} or DEP force\cite{moon2009dielectrophoretic} for such real time detection of microorganism were reported, however developing active, reliable and rapid sampling techniques is still an active research problem\cite{mainelis2020bioaerosol}.
Pharmaceutical aerosol drug delivery is another rapidly growing field in aerosol engineering, where the drug loaded on an excipient constitute a particulate of size about 5 micron\cite{dolovich2011aerosol,chan2006dry}, is delivered through inhalation route to the patient's airways or lungs. Density of such synthetic particulate drug may range from that of typical airborne particles to as low as $10kg/m^3$ with sizes of 5 to 30 micron for ultra light \textit{large porous hollow particles}\cite{gharse2016large}. Drug delivery to the lungs through the airways require precise control over the particulate flight based on their particle size, density and flow parameters, to avoid their deposition in the airways\cite{dolovich2011aerosol,pleasants2018aerosol}. Hence \textit{active targeted pulmonary delivery} techniques are being developed to use noninvasive magnetic forces on the particles to drive them precisely to the point of delivery in the lungs \cite{chan2006dry,babincova2009magnetic}. Such manipulation in the presence of significant air flow drag and finite particle inertia through the intricate geometries of airways and lungs could be both complicated and interesting for fluid mechanists \cite{babincova2009magnetic,pourmehran2015simulation}. Mathematical formalism for the analysis of particle movement with magnetic force closely resembles with that of electrostatic forces on particles considered in the present study. 

In spite of such wider context of interest on the combined influence of air circulations and electrostatic forces, fundamental understanding of their interplay on particle dynamics and collection behavior is more complicated than that of diffusion, convection and sedimentation effects\cite{russel1991colloidal,pich2017gas}, because of its dependence on the relative orientation of flow and electric fields.
Hence in this report a generic geometry of two convex electrode surfaces is considered with two configurations of well defined numerically fabricated vortical flow circulations attached to their surfaces for a fundamental study. Particles are introduced in this domain and their trajectories are analysed to understand how the combined influence of air circulations and electric field differ from their individual influence on fine micron sized and larger particles.
From which, insights on the above discussed aspects of the three aerosol applications are derived; They are, a. how air circulations orientation and strength influences the particulate capture, b. how airborne matter of a certain size may selectively be separated and collected and c. how an external non invasive manipulation influences air circulations between narrow surfaces. 
\section{\label{sec:problem} Problem setup of flow and electric fields}
\begin{figure}[!htbp]
\includegraphics[width=\textwidth]{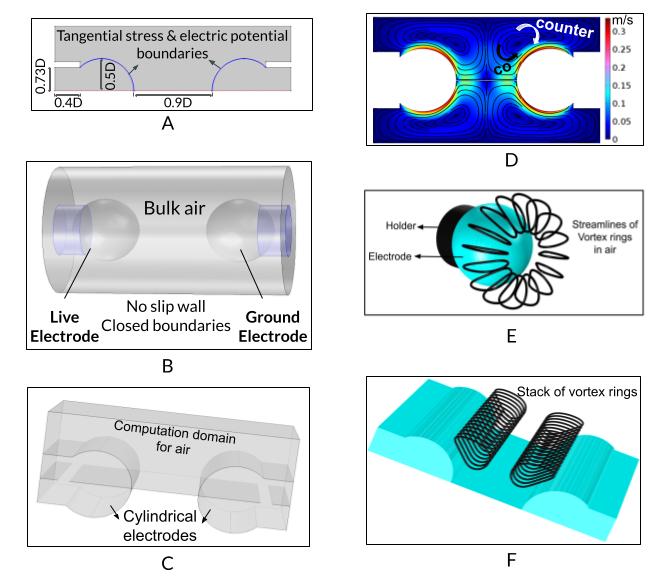}
\caption[Numerical problem setup of vortex circulations subject to electric field]{\label{fig:geometry,vortices} Numerical problem setup (A).2D simulation domain, with boundaries where tangential stress and constant electric potential are applied. $D/2$ is the radius of curvature of the two curved boundaries. (B,C). 3D view of a pair of spherical and cylindrical electrodes attained by revolving and extruding the 2D simulation geometry, respectively. (D). Surface colour and streamlines indicate the velocity field in air, in either of the two orientations of the vortices directions with respect to the attached electrode curvature; Marked as co-direction vortices from pole to equator (black) and counter-direction vortices from equator to pole (White). (E,F). Streamlines rendered in 3D indicate torus and stack of vortex rings between the pair of spherical and cylindrical electrode surfaces, respectively.}
\end{figure}
Numerical setup of flow and electric fields are prepared, to maximize the span of a steady vortex circulation on the curved electrode surface for subsequent particle trajectories analysis in the domain. A geometry with two curved arcs, of diameter $D$ as shown in figure \ref{fig:geometry,vortices}A, represent a generalized pair of convex surfaces in air domain. This geometry can be revolved in to a pair of spheres or extruded out of its plane in to a pair of cylinders of electrodes in air, as shown in figure \ref{fig:geometry,vortices}B\&C respectively. Navier-Stokes equations describing the air flow are\begin{align}\label{Eq:NS}
\rho_a \nabla.\textbf{v} &= 0 \\
\rho_a (\textbf{v}.\nabla)\textbf{v} &= \nabla.(-p\textbf{I}+\mu_a (\nabla \textbf{v}+(\nabla \textbf{v})^T))\end{align} solved in the air domain. $\rho_a$ and $\mu_a$ are the density and viscosity of air. To create steady well defined configuration of vortices of air circulations on the electrode surfaces (like a text book problem), a tangential boundary force $F_t$ is applied on air at the curved boundary arcs (surfaces in 3D) of electrodes. \begin{align}\label{Eq:boundarystressCondition}
(-p\textbf{I}+\mu_a (\nabla \textbf{v}+(\nabla \textbf{v})^T)).\hat{\textbf{n}}&= F_t\hat{\textbf{t}}
\end{align}This is similar to how Hill's solution to vortex rings around droplets\cite{norbury1973family, scase2018hill,chapman1967formation, chen2003transient} invokes a tangential stress continuity on the interface between the two fluids\cite{batchelor2000introduction,acheson1991elementary}. No slip condition is applied on the remaining boundaries of the computational domain. Here, $\hat{\textbf{n}}$ and $\hat{\textbf{t}}$ are unit vectors along outward normal and tangent of the electrode boundary, respectively. Magnitude of $F_t$ is adjusted  between $0.02 - 0.09 N/m^2$ to create vortices of highest velocity magnitude close to electrode of $0.1-0.45m/s$. Direction of circulation is either along equator to the pole of the curved electrode or reverse depending on the direction of tangential stress, identified as \textit{counter} and \textit{co circulation} directions respectively, as shown in figure \ref{fig:geometry,vortices}D. Electrostatic Poisson equation is solved in the air domain with constant electric potential boundary conditions on the electrode boundaries.\begin{align}\label{Eq:Poisson}
\textbf{E}  &= - \nabla V\\
\epsilon_0 \epsilon_a \nabla.\textbf{E} &= \rho_v
\end{align}These equations are solved using finite element methods package COMSOL Multiphysics, AB/COMSOL, Inc. 

\begin{figure}[!htbp]
\includegraphics[width=\textwidth]{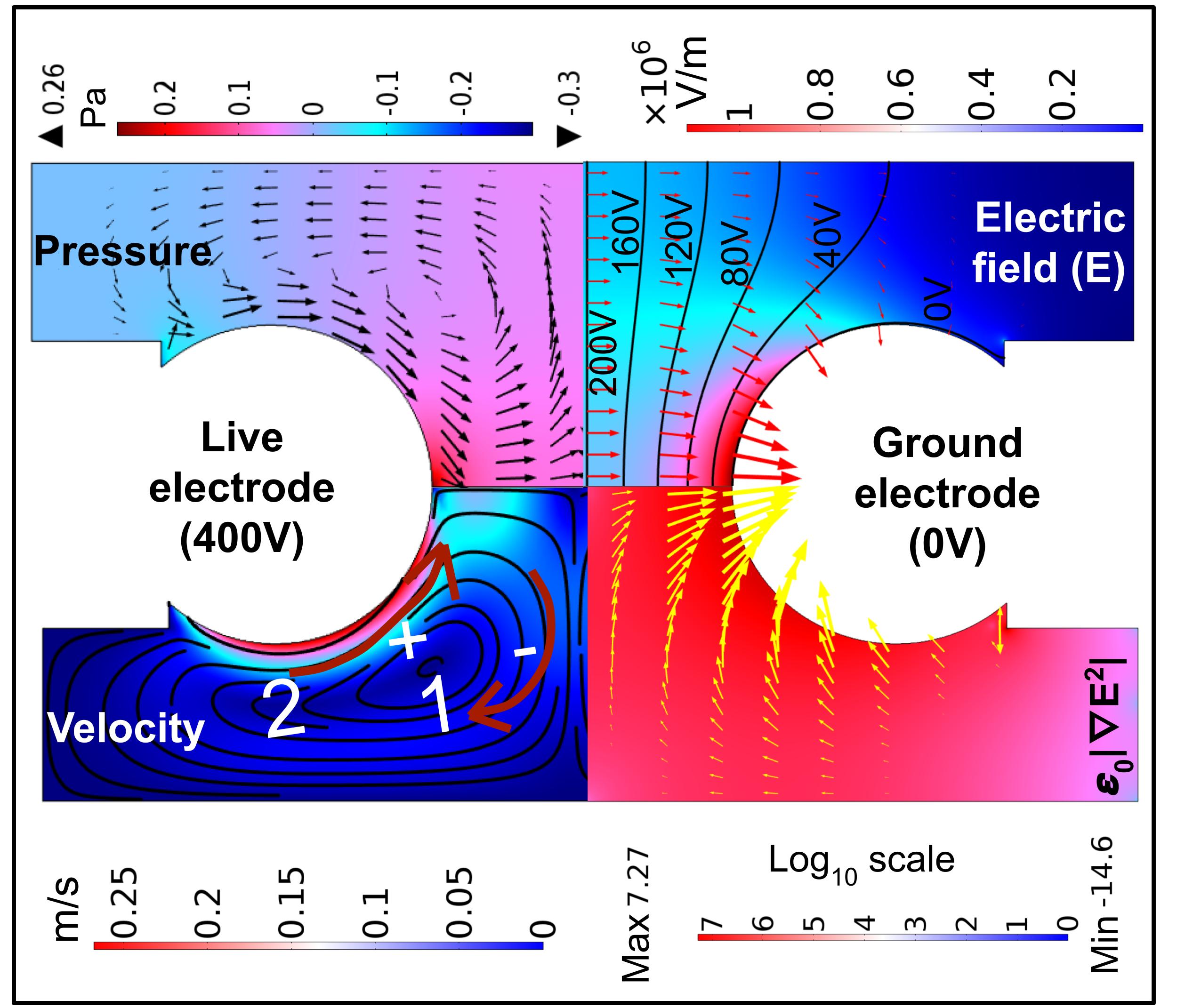}
\caption[2D surface plots of vortex air flow and electric fields]{\label{fig:surfaceplots} Counter vortices between active electrodes. All values in SI units. [Upper left tile:] Colour-pressure; arrows- velocity vectors. [Lower left tile:] Colour-velocity magnitude; Streamlines are in black with segment 1 \& 2 of positive(+) and negative(-) local direction of curvature as marked in red. [Upper right tile:] Colour and red arrows indicate electric field ($|\textbf{E}|$) strength and vectors respectively. Contours of electric potential are shown with labels in black. [Lower right tile:] Colour \& arrows indicating magnitude on log scale \& vector directions of $\epsilon_0 \nabla(|\textbf{E}|^2)$ respectively, representative of spatial variation of dielectrophoretic (DEP) force field on particulate.}
\end{figure}

2D vortices thus created (figure \ref{fig:geometry,vortices}D) forms a torus of vortex rings around a pair of axisymmetrically placed spheres upon revolving and a stack of vortices around a pair of semi-cylinders upon extrusion perpendicular to the surface, as shown in figure\ref{fig:geometry,vortices}E\&F respectively. Similar vortices may be created from mechanically induced flows, boundary layer separation or controlled ion injection (instead of the whole air ionization), as demonstrated in the appendix \ref{append:vortexCreationMethods}. However the source and creation of the circulations is out of scope of current study. Circulation of air in the vortices is unidirectional in either counter or co direction as mentioned above. On the other hand, since fluid path lines are curled around the surface of the electrodes, each streamline is curled to have two segments, segment 1 (marked as negative(-)) and segment 2, which is close to the electrode surface (marked as positive(+)), as shown in the figure\ref{fig:surfaceplots}. Their local radius of curvatures are opposite to each other, towards and away from the vortex center, respectively. Such distortion of the vortices are common in flows between convex surfaces, which is recognized to have a profound influence on airborne particulate dynamics in this study. 

The electric field and gradient of electric field in air in figure \ref{fig:surfaceplots}, indicate how Coulombic and dielectrophoretic electrostatic force fields vary, respectively. Nominal electric field in the domain is of scale $E_0 = \frac{V_0}{D}\sim 10kV/cm$. DEP force on a particle is dependent only on the gradient of electric field and hence symmetric in the domain about a line (or plane) of symmetry between the two oppositely polarized electrode surfaces. However, the direction of Coulombic force is from one electrode to the other depending on the polarity of charge on the particle. Figure \ref{fig:surfaceplots} indicates a counter direction vortex flow circulation which opposes the DEP force on a particle in front of the electrodes;

\section{\label{sec:method} Methodology of particle tracking}
\subsection{\label{subsec:methodNumericalParticles} Complete numerical analysis}
\begin{table}
    \centering
    \begin{tabular}{|c|c|c|}
\hline
Relative permittivity of particle & $\varepsilon_{p}$  & 10 \\ 
\hline
Electrical conductivity of particle & $\sigma_{p}$  & $10^{-4} S/m$ \\ 
\hline
Diameter of particles & $d_p$ & $0.1-50\mu m$\\
\hline
Density of particles & $\rho_p$ & $120-1200 kg/m^3 $ \cite{khillare2012airborne,gharse2016large}\\
\hline
Radius of curvature of electrodes  & $R_E$ & $0.15-0.5mm$ \\ 
\hline
Applied potential difference & $V_0$ & $0-5000V$ \\ 
\hline
Maximum velocity of air in vortex & $v_0$ & $0.1-0.45m/s$ \\ 
\hline
\end{tabular}
\caption{Range of material properties of airborne particles, geometrical and operating parameters considered for the study.}
\label{tab:part_prop}
\end{table}
Nearly spherical airborne particles of density $\rho_p$ and aerodynamic diameter $d_p$ with no or mild natural charge $q_p=z_p e$, with a range of properties indicated in table \ref{tab:part_prop}, are introduced into the computation domain under a range of operating conditions; where $e$ is the charge of an electron and $z_p$ is the total charge number. Time dependent position $x_p$ of each Lagrangian particle of mass $m_p$ under the influence of Stokes drag from the vortical air flow ($f_d$), gravity ($f_g$), dielectrophoretic (DEP) force ($f_{dep}$) and Coulombic electrostatic force ($f_{C}$) are numerically estimated from particle momentum conservation equation, 
\begin{align}\label{Eq:ParticleMomentum}
   m_p \pdv[2]{\textbf{x}_p}{t} &= {\textbf f}_{d}+\textbf{f}_g+{\textbf f}_{dep}+ \textbf{f}_{C}\\
    {\textbf f_d} &= \frac{m_p}{\tau_p}({\textbf v}-{\textbf v}_p) =3 \pi \mu_a d_p ({\textbf v}-{\textbf v}_p)\\
\textbf{f}_g &=m_p \textbf{g}\\
    {\textbf f}_{dep}&= \frac{\pi}{4} d_p^3 \epsilon_0 \epsilon_a \kappa_{CM}\nabla|\overline{\textbf{E}}|^2\\
    \textbf{f}_{C}&= q_p \textbf{E}
\end{align}Where, $\tau_p= \frac{\rho_pd_p^2}{18 \mu_a} \sim 0.1ms$ is the viscous drag relaxation time scale of particles, $\textbf{v}$ is the velocity of particle and  $\kappa_{CM} = \frac{\epsilon_p-\epsilon_f}{\epsilon_p+2\epsilon_f}\sim 1$ is the Clausius Mossoti factor respectively. Reynolds number of the air flow at the vortex length scale $Re=\frac{\rho_a v_0 D}{\mu_a} \sim 25$ and particle length scale $Re_p=\frac{\rho_a v_0 d_p}{\mu_a} \sim 1$ both are low, hence Stokes formulation is used for drag force on particles\cite{hinds1999aerosol}. Particle acceleration effects such as Basset force and added mass effects, are ignored in the steady velocity field. 

Dimensionless form of this equation with length and velocity in the equations scaled with electrode diameter $D$ and velocity scale $v_0$ respectively and electric field scaled with $E_0 = \frac{V_0}{D}$, where $V_0$ is the potential difference between the electrodes, is\begin{align}\label{eq:dimless_full_particle_momentum}
    St_v \pdv{\tilde{\textbf{v}}_p}{t} &= - ({\tilde{\textbf{v}}}_p-{\tilde{\textbf{v}}}) +St_g \tilde{\textbf{g}}+ n_{D} {\nabla} |{\tilde{\textbf{E}}}|^2 + n_{C}{\tilde{\textbf{E}}}
\end{align}Where, vortex Stokes number $St_v = \frac{\tau_p}{\tau_0}= \frac{\rho_p d_p^2}{18 \mu_a} \frac{v_0}{D}$, the ratio of particle relaxation time scale under Stokes drag $\tau_p \sim 1\mu s - 10ms$ and vortex air circulations time scale $\tau_0=D/v_0 \sim 1 ms$ represents the relative magnitude of particle inertia in the vortex with respect to the aerodynamic drag. $St_g=\frac{\tau_p}{\tau_g}=\frac{g_0}{v_0 \tau_0}\lesssim 0.1$ represents gravity is weaker to aerodynamic drag. Similarly, fig \ref{fig:DEP_dragForces} in appendix indicate gravity force $f_g$ is much weaker to electrostatic forces for the whole range of particle size and density, hence ignored in the computation. Dimensional form of the Lagrangian equations are solved and numerical convergence of results is ensured. $n_{D} = \frac{\epsilon_0 \epsilon_a \kappa_{CM}}{12 \mu_a} \frac{d_p^2 E_0^2}{D v_0}$ and $n_{C} = \frac{e E_0 z_p}{3 \pi \mu_a d_p v_0}$ are the Dielectrophoretic number and Coulombic force number, representing the strength of DEP and Coulombic electrostatic forces on the particles relative to the drag force on them, respectively. For the parameter range of the study, in the table \ref{tab:part_prop},  $n_{D}$ and $St_v$ of this system range between $10^{-6}-10^{1}$. Ratio of these numbers,  $\xi_v = n_D/St_v$ is the relative magnitude of inertial effects over DEP force, which represents the operating conditions in the combination of significant electric and vortex flow fields compared to viscous drag effect. Hence, the two scaled parameters $\xi_v$ and scaled particle size $d_p/D$ exclusively and exhaustively represents dynamics of particles in the vortices subject to electric field beyond the viscous dominance limit. 

An asymptotic steady state of particle dynamics arises eventually at a time scale of $t_0$ recognized as around $100ms$, as discussed in the results, figure \ref{fig:captureQualitative}. Particles which contact the electrodes are stick to the electrodes and counted as \textit{particles collected}. Particles contacting any other spatial boundaries of the computational domain are frozen on them from any further movement and considered to be \textit{deposited on the walls}. Remaining are counted as airborne. Fraction of particles in each of these states at the steady state $N_C,N_D,N_A$ respectively, signifies the probability of state of particle at steady state, with an arbitrary initial location in the domain.
\subsection{\label{subsec:MethodCrescent} Particles in a crescent model vortex}
\begin{figure}[!htbp]
\includegraphics[width=\textwidth]{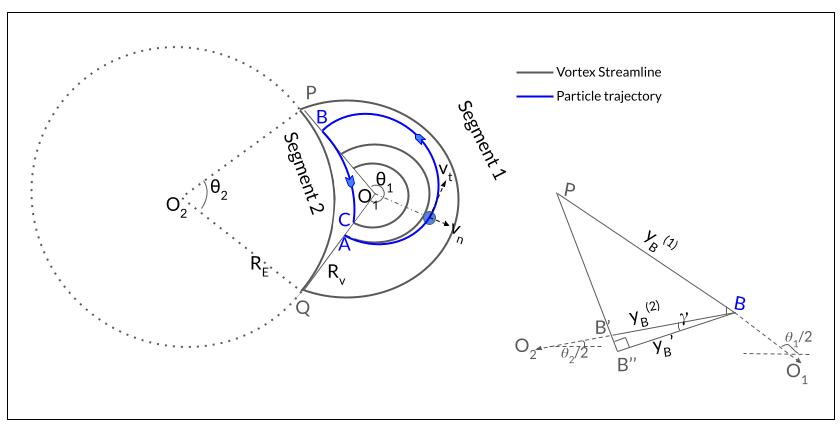}
\caption[Illustration of crescent model vortex]{\label{fig:crescent_model} Illustration of crescent model vortex comprising of two circular segments on each of the streamlines, shown in brown. Particles movement A-B-C in the vortex is shown in blue, illustrating inertial deviation of particle trajectory away from and towards the vortex center $O_1$ in segment 1 \& 2 respectively. Insert on the right side of the figure indicates, geometrical setup of conversion of distances from point $B$ to $O_1$ and $O_2$, as described in appendix \ref{append:subsec:CrescentModel}.}
\end{figure}

A streamline on the vortex shown in figure \ref{fig:surfaceplots}
can be approximated to be a crescent comprising of two circular arcs, as shown in figure \ref{fig:crescent_model}. Segment-1 indicates the region of counterclockwise angular velocity and segment-2 indicates that with clockwise angular velocity around their centers $O_1, O_2$ respectively. Segment-1 consists of concentric circular arcs of a constant angular width $\theta_1$while the angular width of concentric circular arc streamlines on the other segment varies suitably. 
$R_v$, the radius of  largest segment 1 of streamlines is considered as the outer boundary of the vortex domain. $R_E=\frac{D}{2}$, the radius of curvature of the electrode surface is the minimum radius of curvature of segment 2 streamlines. Angular velocity on each of the segments is considered uniform and equal to $\omega_1$ and $\omega_2$ respectively. 

This vortex model is used to estimate trajectory of a particle with dominance of aerodynamic drag and inertial influences among all. Then movement of a particle within each of the circular segments can be estimated as, 
\begin{align}
  r_{p(\theta_p)} &= r_{p(0)} e^{\theta_p/\omega_i \tau_{pn}}
\end{align}
Where, $r_{p(\theta_p)}, \theta_p$ are the radial and angular coordinates of the particle with initial position $r_{p(0)}$, with respect to the corresponding center of curvature of the segment and the axis $\overline{O_1O_2}$. $\tau_{pn}= \frac{2 St_v}{\sqrt{1+4St_v^2 \omega_v^2}-1}$ is the time scale of particle radial migration in the circular vortex segment. Particle movement in a counter vortex under electric field may be represented as follows. Consider a particle with initial position $A$ in segment 1 at the boundary of the segments, moves to points $B \& C$ as it jumps between the segments, as shown in the figure \ref{fig:crescent_model}. Distance and angle from the center of segment 1 ($O_1$) and that of segment 2 ($O_2$) are interchangeable according to the geometric arrangement illustrated for position $B$, in the figure \ref{fig:crescent_model} and derived algebraic expressions in the appendix \ref{append:subsec:CrescentModel}. Hence, coordinates of these points $B\&C$ with respect to $O_1$under weak electrostatic forces can be estimated as derived in the appendix \ref{append:subsec:CrescentModel},
\begin{align}\label{eq:rb_in_ra}
    r_{p(B)} &= r_{p(A)} e^{\frac{\theta_1}{\omega_1 \tau_{pn1}}}
\end{align}
\begin{align}\label{eq:rc_in_ra}
r_{p(C)} &= r_{p(A)} \left( e^{\frac{\theta_1}{\omega_1 \tau_{pn1}}+\frac{\theta_2}{\omega_2 \tau_{pn2}}} \right) + \left[ R_v \left(1- e^{\frac{\theta_2}{\omega_2 \tau_{pn2}}} \right) + R_D \left( \frac{1}{\sqrt{2}}-e^{\frac{\theta_2}{\omega_2 \tau_{pn2}}} \right)\right] 
\end{align}Radial widths of segments and air velocity values in the crescent vortex are matched with the CFD results, to study the effect of vortex shape on the airborne trajectories of particles.
\section{\label{sec:results}Results}
\subsection{\label{subsec:result_traj} Effect of vortex direction on uncharged particles trajectories}
\begin{figure}[!htbp]
\includegraphics[width=\textwidth]{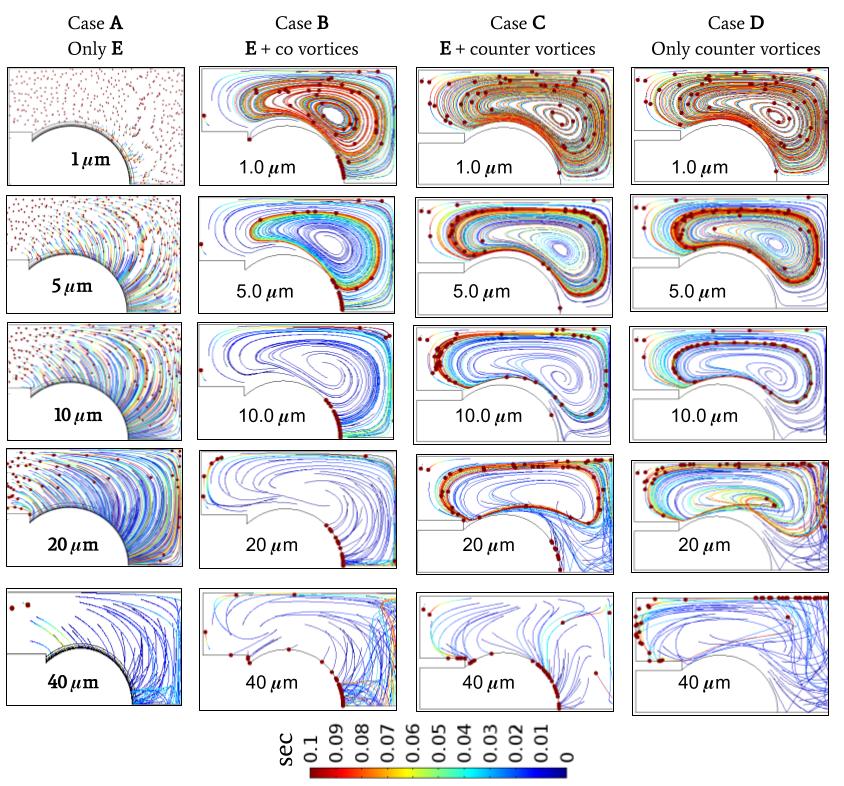}
\caption[Effect of vortex direction on uncharged particles trajectories]{\label{fig:Trajwith,withoutVortex} Comparison of trajectories of uncharged particles subject to, Case-A only electric field, Case-B co-vortices and electric field, Case-C counter vortices and electric field and Case-D only counter vortices circulations. Electric potential difference and vortex velocity maximum used are $400V$ and $0.34m/s$ respectively, corresponding to $\xi = 0.046$ for cases B\&C with $St_v$ and $n_D$ varying in the range of $0.003 - 4$ and $0.0001 - 0.2$ respectively. 
Trajectories of particles (indicated as dots) of different sizes (marked on each tile of image) in one half of the symmetric 2D domain of each case are shown. Color legend of the trajectories indicate the duration of particle flight from its initial position to reach that location. Constructive and antagonist effects of vortices with the DEP particle capture onto electrodes can be noticed, with co and counter circulations of case B \& C respectively.}
\end{figure} 
Trajectories of uncharged particles of density $1200 kg/m^3$ with different sizes, $0.1s$ after their release from arbitrary locations between the electrodes of potential difference 400V are shown in figure \ref{fig:Trajwith,withoutVortex}, in one half of the symmetric domain. Similar results were achieved by using Schiller Naumann drag formula subject to the finite Reynolds number of the flow, ensuring correctness in the usage of Stokes drag formulations. Case A represents when the ambient air is stagnant and particles are attracted towards the electrode front surfaces due to DEP. DEP force influence visibly increases with increasing size of particles. Case B \& C represent a significant influence of co-vortices and counter vortices of air circulations on the trajectories of particles respectively, especially on particles of size below $20\mu m$. Electric field between the electrodes is same as case A. Co-vortices brings the smaller particles close to the electrode front surfaces where DEP is higher, hence visibly improve their electrostatic capture on electrode.

On the other hand the counter vortices in case C, trap the $1\mu m$ particles circulating in the air streamlines everywhere in the domain and $5-20\mu m$ particles in rings of airborne trajectories; Thus resulting in a visible decrease in their capture on electrodes. These trajectories are similar to case D where electric field is switched off on these counter vortices circulation. DEP influence dominates on larger particles of size $40 \mu m$ and above, in case C. In the absence of electric field in case D these large  particles can be noticed to be expelled out of the vortex on to the side walls of domain due to particle inertia. 
\subsection{\label{subsec:result_capture}Effect of vortex direction on particle capture}
\begin{figure}[!htbp]
\includegraphics[width=\textwidth]{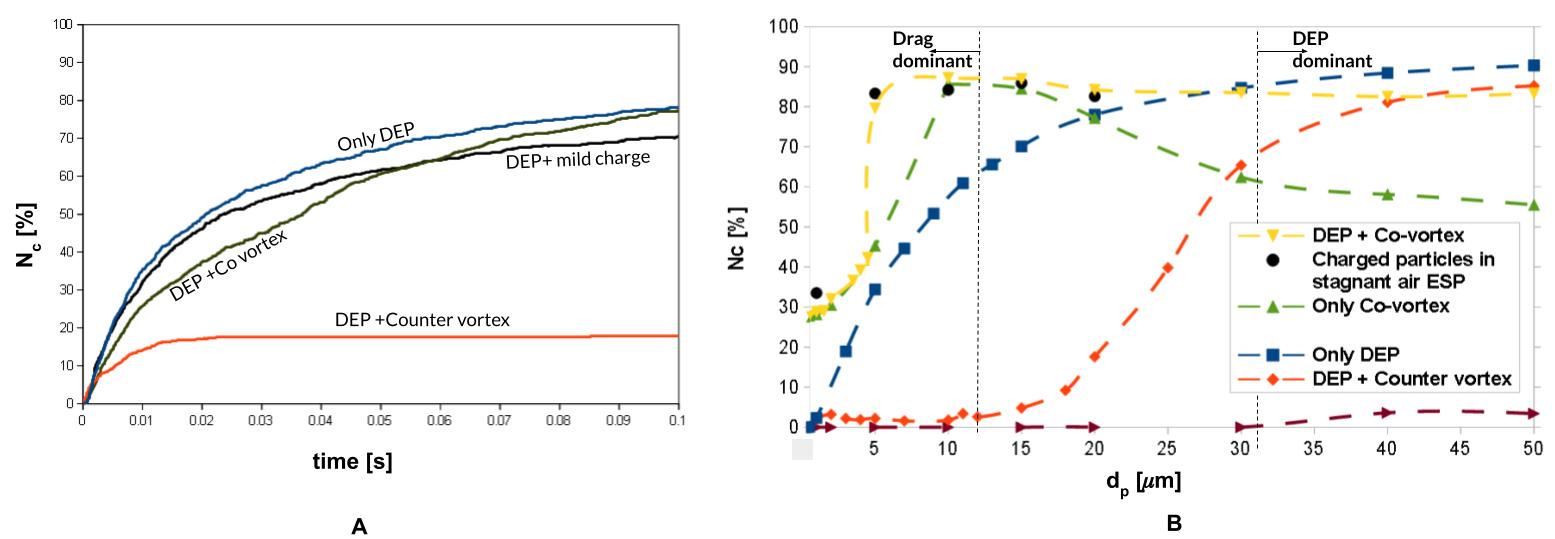}
\caption[Influence of flow circulations on capture of uncharged particles on the electrodes.]{\label{fig:captureQualitative} (A) Fraction of particles (of density $1200kg/m^3$) collected on electrode surfaces in the closed air domain for different operating cases, indicating an asymptotic steady state after 0.1sec. Mild charge is equivalent to the definition in section \ref{subsec:anal_charge}, while other cases are uncharged particles. (B) Different curves indicate influence of flow circulation configuration and electrostatic field on the particle capture on electrode surfaces. Maximum velocity of air flow is 0.34m/s and electric potential difference between electrodes is 400V in these figures, combination of these manifests as $\xi =0.046$. Black dots indicates equivalence of the case of co-vortices subject to electric field with highly charged particles in the electric field in a stagnant air. Such equivalent charges are $25,880,1500,2500,3650$ elementary negative charges on particles of sizes $1\mu m,5\mu m,10\mu m,15\mu m,20\mu m$ respectively.}
\end{figure}
Figure \ref{fig:captureQualitative}.A indicates a visibly exponential rise in particle capture probability with time and an asymptotic steady state of particulate capture dynamics at about 100ms, for different cases of air circulations subject to electric field. It is interesting to notice that, particles with mild charge estimated as a linear function of their size, $q_p=-\left(4 \left( \frac{d_p-0.3[\mu m]}{9.7\mu m}\right)+1\right)$ have a slower capture than the uncharged, due to the unipolarity of charges which will be elaborated later in section \ref{subsec:anal_charge}. Figure \ref{fig:captureQualitative}.B indicates the capture probability of particle of different sizes with density $1200kg/m^3$ and electrical properties as mentioned in table \ref{tab:part_prop}, under different operating cases of vortices circulations directions  and applied electric field. Co-vortices without any electric field demonstrate higher particle capture than the pure DEP capture of particles below $20 \mu m$, which is the typical size range of aerodynamically influenced \textit{fine particles} in ESPs\cite{lu2017analysis}. Above this particle size, electrode collection lowers due to inertial spillage on to domain walls. However combining these vortices with the electric field improves the particle capture through-out the wider range of particle sizes. This collection efficiency of uncharged particles is comparable to that of particles carrying high charges of up to thousands of electrons in a stagnant air ESP, as shown in the figure  \ref{fig:captureQualitative}.A and its caption.

On the other hand, counter vortices without any electric field demonstrates negligible ability to deliver the particles on to the electrode surface. When they are combined with the electric field, vortices restrict the particle capture for sizes up to a critical diameter $d_{p0}$, the \textit{cut-off size}, that is about $15\mu m$ for the case demonstrated in the figure \ref{fig:captureQualitative}. However rise in capture count with particle size is similar to that in case of DEP in stagnant air. Hence while capture improves monotonously with the co-vortices, it needs further investigation in case of counter vortex subject to electric field.
\subsection{\label{subsec:result_modes}Particle states in different operating regimes of counter vortices under electric field}
\begin{figure}[!htbp]
\includegraphics[width=\textwidth]{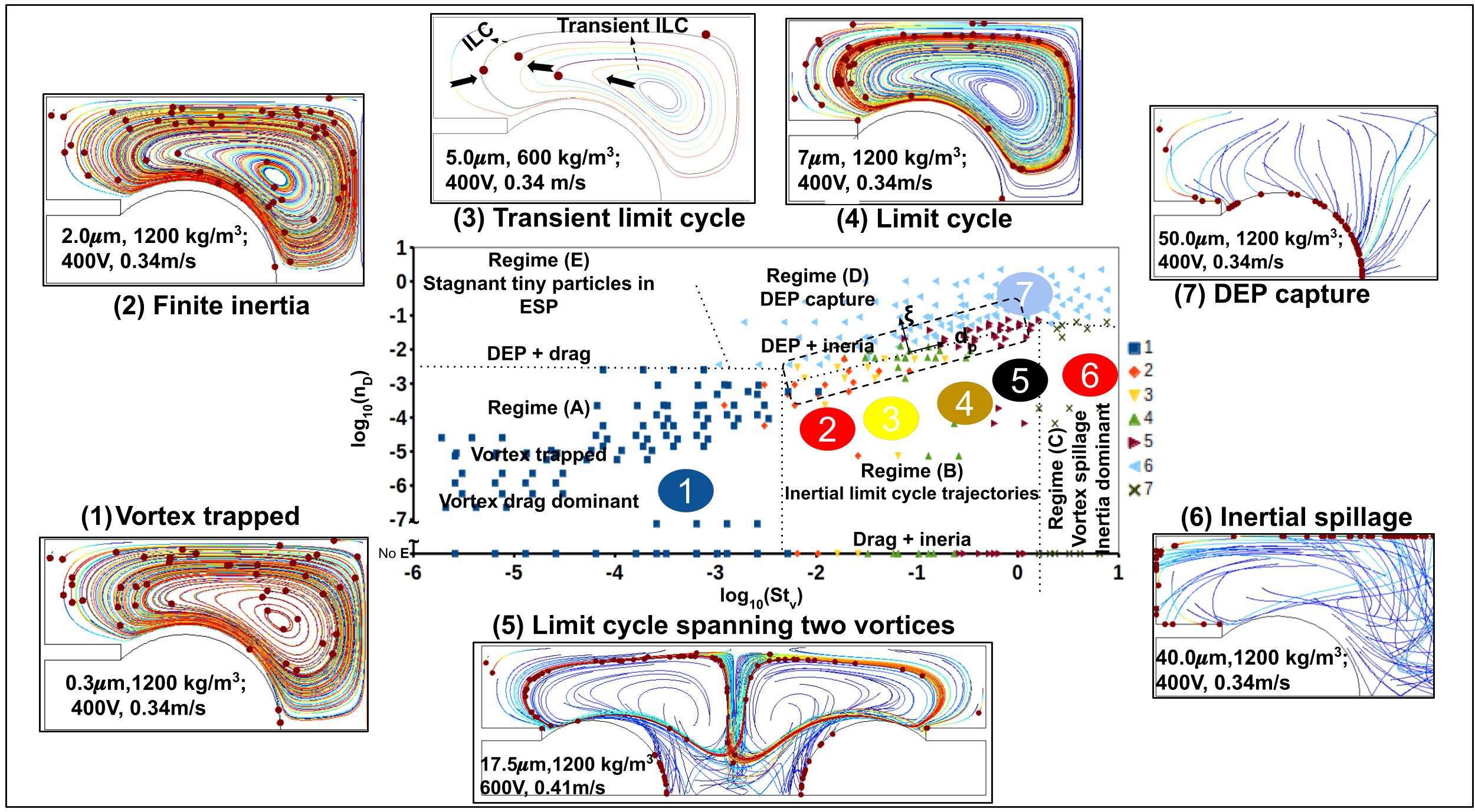}
\caption[Various modes of particle movements.]{\label{fig:ModesPraticleTraj} Various regimes of relative dominance of drag, inertia and DEP on uncharged particle trajectories mapped on the dimensionless numbers $n_D$ and $St_v$. Cases from a range of parameters from table \ref{tab:part_prop} are used. Snippets indicate representative case of particle trajectories for different states of particle behaviors. }
\end{figure}

Drag force on the particles scales with size of the particle $d_p$, while the dielectrophoretic (DEP) electrostatic force and inertial effects of the particle scales with $d_p^3$. Hence as the particle size increases, DEP and inertia dominates. While the DEP, depending on the particle relative permittivity $\epsilon_p$, attracts the particles towards the electrodes, the inertial effects based on particle density drifts them along the local tangent of their trajectories. On the other hand, vortex drag deviates the particles from the electrostatic trajectories and circulate them with the air flow. Seven (7) distinguishable states of particle behavior in the counter vortices subject to electric field are mapped on $(St_v,n_D)$ coordinates on log scales, as shown in the figure  \ref{fig:ModesPraticleTraj}. These dimensionless numbers may be interpreted as a represent of combination of dimensional parameters particle size, air velocity in vortices and electric field strength. For example, for particles of density $1200kg/m^3$ and electrical properties from table \ref{tab:part_prop}, $n_D= 3\times 10^{-8} E_0^2 d_p^2/v_0$ and $St_v=3.8 \times 10^{-6} d_p^2 v_0$, with all variables in SI units. Various regimes of operating situations are identified as shown in this phase diagram, such as an extension of the phase diagram of diffusion, drag convection, inertia and gravitational sedimentation of particles described by Saville et.al.\cite{russel1991colloidal} with an addition of electrostatic forces.

Left-bottom corner of this phase diagram represents the aerodynamic drag dominant regime. Influence of inertia and DEP on particles increases along the X \& Y axis directions respectively. In \textit{Regime A} vortex drag dominates and traps the particles to circulate with the air streamlines, that is typical of submicron or order $1\mu m$ size of ambient uncharged particulate in the vortices with low applied electric field gradients. Between regime A and the inertia dominant particle spillage \textit{regime C} (on right bottom corner) different states of particle flight trajectories are identified in the the drag-inertia transient \textit{regime B}, as shown in the snippets $2-5$ of the figure. With a mild influence of \textit{particle finite inertia} they start to deviate from the air streamlines indicated as particle state 2. As the inertia increases further, particles move into stable rings of limit cycle trajectories. Particles located either outside or inside such trajectories move towards these trajectories. Particles which need longer time to completely get into these trajectories are identified as the state 3, \textit{transience of limit cycle} formation. On further increase in inertia, particles fall into the \textit{limit cycles} well within the time scale of the analysis, presented as state 4. Upon further increase in particle inertia the stable \textit{limit cycle trajectories span across multiple vortices}, i.e., particles periodically jumps between the vortices, identified as state 5. For the typical properties of common airborne particles DEP in this configuration becomes significant, at $E_0 \sim 1-10kV/cm$ in this regime B and thus distorts the limit cycle shapes which in turn influences the particle capture on the electrodes, as analyzed in section \ref{subsec:anal_ILC}. 

 The border region between the regime B and the DEP dominant strong vortex \textit{regime D} is of fundamental interest to understand the intersection of drag, inertial and DEP all the three influences on the particles. This regime can also be addressed in an alternative set of coordinates $(\xi_v=\frac{n_D}{St_v},\hat{d}_{p}=\frac{d_p}{D})$ which are independent of each other, as discussed earlier in section \ref{subsec:methodNumericalParticles}. Analysis of how particle capture varies in the transition from regime B to D is analyzed in detail, in section \ref{subsec:anal_capture}. Regime E on the left upper corner represents low significance of air circulations hence resembles pure electrostatic precipitators with stagnant air.
\section{\label{sec:Analysis}Analysis}
\subsection{\label{subsec:anal_capture}Inertial cutoff of electrostatic particle capture in counter vortices}
\begin{figure}[!htbp]
\includegraphics[width=\textwidth]{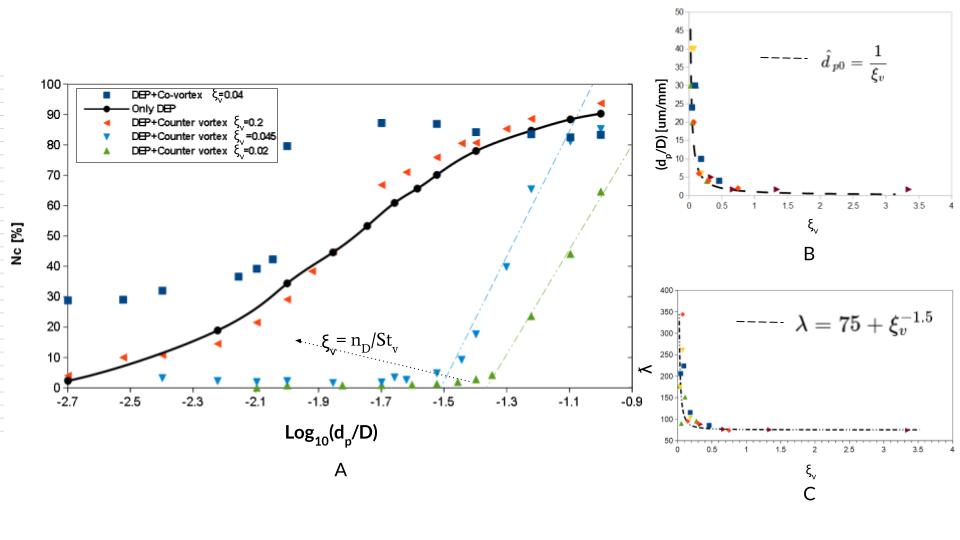}
\caption[Particle capture in counter vortices.]{\label{fig:counterVort_capture} (A) Particle capture in co-vortices and counter vortices with different values of $\xi_v$ are compared with no air flow case, all under a same electric potential difference between the electrodes. A linear-log relation of the particle capture with scaled particle size is noticed with intercept corresponding to $\hat{d}_{p0 (\xi_v)}$ and slope $\lambda_{(\xi_v)}$. (B) and (C) indicate the variation of   $\hat{d}_{p0 (\xi_v)}$ and  $\lambda_{(\xi_v)}$respectively, with $\xi_v$. Data points of each legend indicate different cases with varying particle densities, nominal electric field, vortex velocity scale and electrode sizes. Functional fits with equations shown are indicated with the dashed lines.}
\end{figure}

Variation of particle capture probability with scaled particle size $\hat{d}_p=\frac{d_p}{D}$ in the transition from regime B to D of counter vortices is compared with that of co-vortices and linear log variation relation of stagnant air DEP cases, in figure \ref{fig:counterVort_capture}. X-intercept indicates the particle cut-off size ($\hat{d}_{p0}$) and linear fit slope indicates the capture rise with increase in particle size ($\lambda$), which in case of counter vortices vary based on the combination of flow and electric fields, $\xi_v$. Variation of $\hat{d}_{p0}$ and $\lambda$  with $\xi_v$, for different cases of applied voltage, vortex velocity, electrode radius and particle density and diameter from table \ref{tab:part_prop} are represented in figures \ref{fig:counterVort_capture}B \& C. The empirically fit expressions are  $
\hat{d}_{p0(\xi_v)} = \frac{1}{\xi_v}$ and $\lambda  = \lambda_0 + \xi_v^{-1.5} with  \lambda_0 = 75$. Hence dependence of particle capture probability on particle size and operating conditions can be expressed with a single expression as, 
\begin{align}
N_{c(\hat{d}_p, \xi_v)} [\%]= log_{10}(\xi_v (\hat{d}_p)^{\lambda_0 + \xi_v^{-1.5}})
\end{align}Data shown in the figure \ref{fig:counterVort_capture}B\&C indicates a sudden shift from vortex cutoff to DEP dominant capture around the value $\xi_v \sim 0.15\pm 0.05$, while maintaining the linear log nature of the curve in figure \ref{fig:counterVort_capture}A. Hence the shift in counter flow inertial resistance to the electrostatic particle can be quantified in terms of a \textit{vortex capture deterioration number} $\Xi_v=\frac{1}{\xi_v}= \frac{St_v}{n_D} \sim 5-10$. Which resembles but different from the aerodynamic convection cut-off of inertial capture in traditional Andersen's kind of inertial impactors used for particle collection\cite{andersen1958new,mainelis2020bioaerosol}.
\subsection{\label{subsec:anal_ILC}Inertial limit cycle segregation of particles airborne in counter vortices}
It is of primitive knowledge that inertial limit cycle (ILC) trajectories shown in regime B of phase diagram figure \ref{fig:ModesPraticleTraj}, do not appear for particles in perfect circular vortices. Very few literature is available on inertial limit cycles trapping of suspended particles. It was observed around curved surfaces oscillating at a very high frequency due to viscous streaming flow vortices, by K Chong et.al. both theoretically\cite{chong2013inertial} and numerically\cite{chong2016transport}. Hence the crescent vortex model can be used to confirm that the  distorted shape of the vortex is the reason for this phenomena. Particle movement equations in crescent vortex derived in the section \ref{subsec:MethodCrescent}
can be used in the parameter regime B of figure \ref{fig:ModesPraticleTraj} farther from its boundary with DEP dominant regime D (i.e., when the relative influence of DEP force, $n_D$ is lower). Consider a particle of  initial location $A$ which moves to location $C$ as it completes one complete revolution around the vortex center. From equation \ref{eq:rc_in_ra},
\begin{align}
r_{p(C)} &= \alpha r_{p(A)} +\beta
\end{align}
Where  $\alpha=\left(e^{\frac{\theta_1}{\omega_1 \tau_{pn1}}+ \frac{\theta_2}{\omega_1 \tau_{pn2}}} \right)$ and $\beta =R_v\left(1-e^{\frac{\theta_2}{\omega_2 \tau_{pn2}}} \right) + R_D\left( \frac{1}{\sqrt{2}}- e^{\frac{\theta_2}{\omega_2 \tau_{pn2}}} \right)$. $\tau_{pn1,2} = \frac{2 St_{v1,2}}{\sqrt{1+ St_{v1,2}^2}-1}$.
Limit cycle existence can be identified as, when the net particle migration is zero with the conditions, 
\begin{align}
r_{p(A)}=
\begin{cases}
< r_{p(C)} \quad &; r_{p(A)}< r_{p(LC)}\\
= r_{p(C)} \quad &; r_{p(A)}= r_{p(LC)}\\
> r_{p(C)} \quad &; r_{p(A)}>r_{p(LC)}
\end{cases}
\end{align}
$r_{(LC)}= \frac{\beta}{1-\alpha}$ is the value of $r_{p(A)}$ on the limit cycle trajectory. Since $\alpha>1$ and $\beta<0$ for all geometrical configurations and velocity field magnitudes, $r_{(LC)} >0$ hence limit cycle always exists. However it practically appears only when $r_{(LC)} < R_v$. 

\begin{figure}[!htbp]
\includegraphics[width=\textwidth]{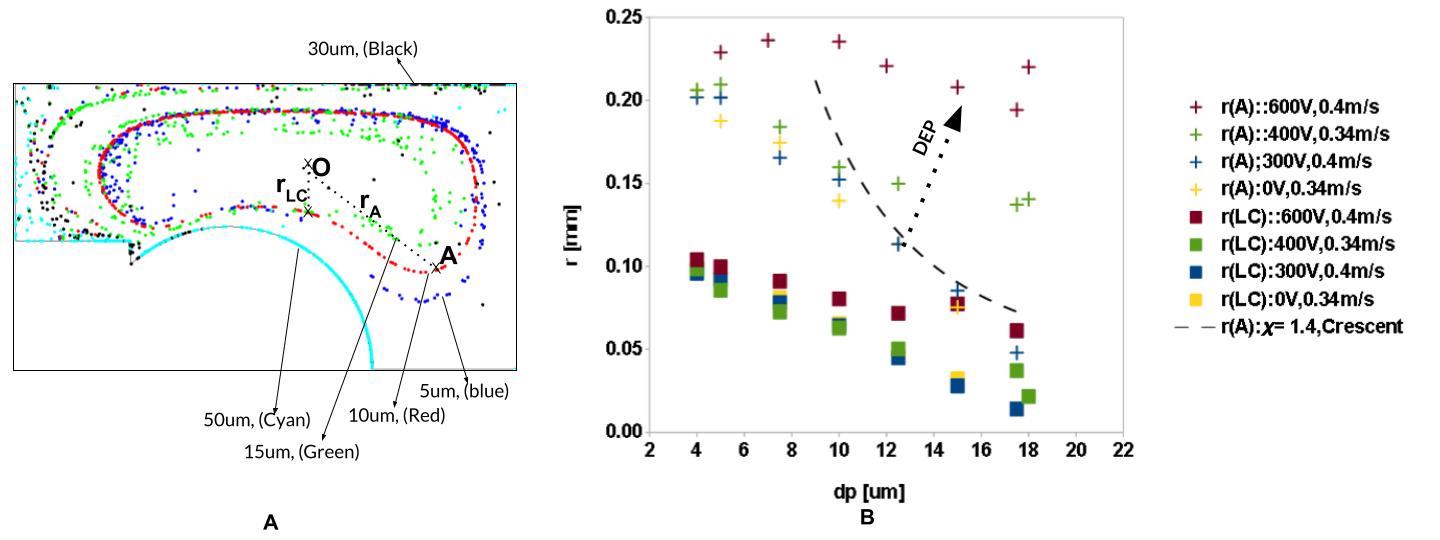}
\caption[Inertial limit cycle trapping and segregation of airborne particles]{\label{fig:ILC} Inertial limit cycle trapping and segregation of airborne particles. (A) Limit cycle trajectories of $5\mu m$, $10\mu m$ and  $15\mu m$ at 400V applied electric potential and 0.34m/s air velocity scale, indicating larger particles are trapped further inside the vortex. $30\mu m$ particles are expelled on to the domain side walls due to inertia and $50\mu m$ particles are electrostatically collected on electrode surface. (B) Variation of ILC trajectory size $r_{(LC)}$ and closest distance between vortex center and the point $A$ (where local direction of particle circular motion changes) $r_{A}$ on the limit cycle trajectory with particle size $d_p$, from the numerical results of complete particle tracking and estimates from crescent model. $\rho_p = 1200kg/m^3, R_v=100\mu m, R_E= 250\mu m$. }
\end{figure}

Figure \ref{fig:ILC}A indicates numerical results of larger or heavier particles stay closer to the vortex center than the smaller or lighter particles trapped in the ILCs. ILC particle trajectories size is indicated as $r_{(LC)}$ and distance $r_{p(A)}$ is used for comparison of crescent model results. Decrease in limit cycle size with increase in particle size of same density is estimated from both the complete numerical study and the crescent vortex model for different values of vortex strengths, as plotted in figure \ref{fig:ILC}B. Particles on segment 2 have higher velocity and radius of curvature than segment 1. The relative influence of these segments in pushing towards and pulling away from the vortex center respectively, can be quantified with the ratio of angular velocities in their respective segments, $\chi = \frac{\omega_1}{\omega_2}$. It is recognized, the trend of ILC size with particle may reverse for different values of $\chi$, hence different spatial distribution of velocity field in the vortex domain, as discussed in the appendix \ref{append:crescentModel}. Such variation may arise from different mechanisms of how the vortical flows originate, such as from the bulk of the domain instead of from the electrode boundary. However ILC trajectories which appear due to the convex shape of the surface as mentioned above, exist in either case independent from such origins of vortical flow circulations. 

As the applied electric field increases, the particles are pulled towards the electrode surface, hence counter balancing the increasing inertial movement into the vortex, resulting in a constant ILC size for the range of particle sizes, as shown in the figure \ref{fig:ILC}A. Thus DEP impedes the airborne segregation of trajectories of particles based on their size and inertia. As the influence of DEP and inertia increases further, both together promotes the particles to jump from one vortex to another resulting in stable ILC trajectories that span across multiple vortices, described as the state 5 in the phase map figure \ref{fig:ModesPraticleTraj}. 
\subsection{\label{subsec:anal_Segregation}Selective deposition of particles for collection}
\begin{figure}[!htbp]
\includegraphics[width=\textwidth]{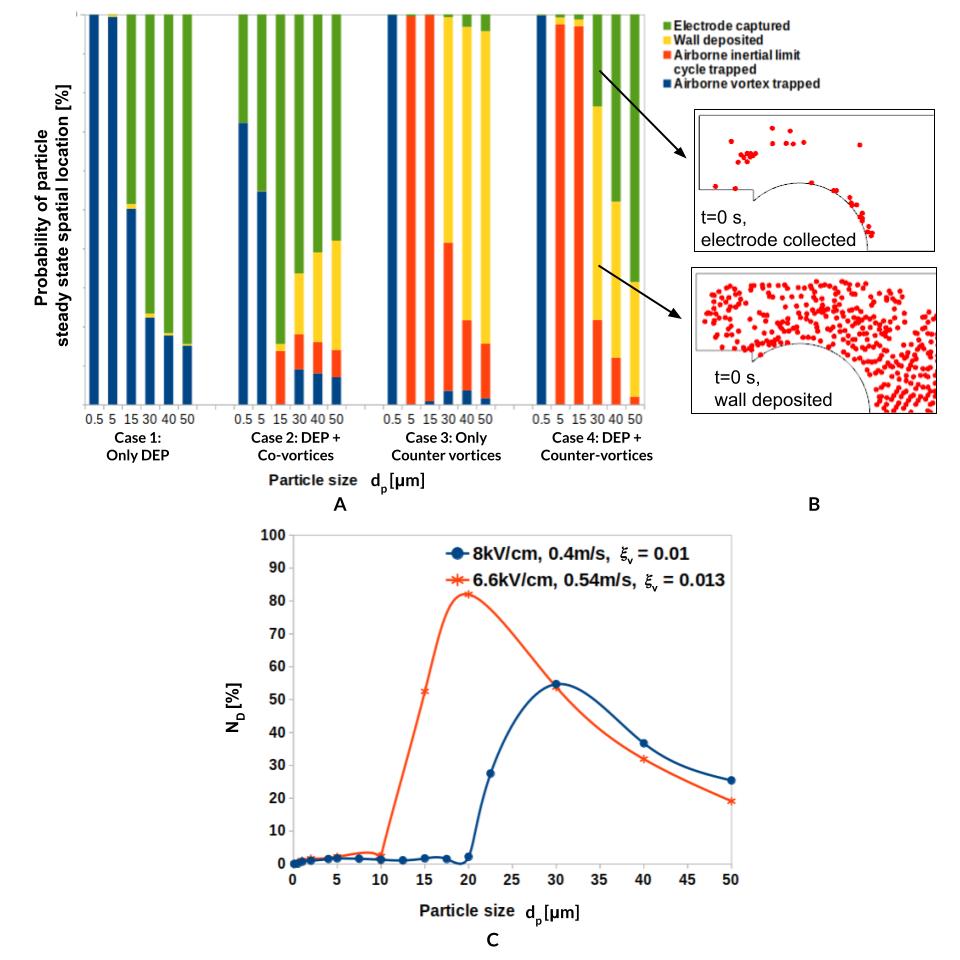}
\caption[Probability of particle spatial location at steady state]{\label{fig:SegregationChart} Probability of steady state spatial location of particles. (A). Particles of density $1200kg/m^3$ in stagnant air (case 1), co-flow subject to electric field (case 2), counter flow vortices without (case 3) and with electric field (case 4). Vortices of maximum velocity in cases 2, 3 \& 4 is 0.4m/s and electric potential applied between the electrodes of diameter $D=0.5mm$ in cases 2 \& 4 is 300V. Probability of finding a particle in different spatial locations is indicated in different colours as labeled. (B). Initial locations of particles which are collected on the electrode or deposited on the side walls by the time of steady state.(C). Probability of finding particles of different sizes, deposited on the side walls at steady state is shown for two operating conditions as labeled.}
\end{figure}

Probability of airborne particle spatial segregation at steady state in different configurations of vortices and electric field are compared in figure\ref{fig:SegregationChart}A, for $300V$ applied between $0.5mm$ diameter electrodes with counter vortices of velocity magnitude $0.4m/s$. Submicron particles are trapped in the air streamlines due to complete dominance of drag force on them. As their size increases, inertial effects in combination with drag, traps the particles airborne in ienrtial limit cycle trajectories. On further increase in particles size and when there is no electric field (in case 2 \&3), larger particles are inertially deposited on electrodes by co-vortices and domain walls by counter vortices. However when electric field is applied on the counter vortices, particles in a size window are deposited on the side walls of the domain, while further larger particles are electrostatically collected on the electrodes. Fig \ref{fig:SegregationChart}B indicates, the initial locations of $30\mu m$ particles which are collected on the electrodes and deposited on the walls by the time of steady state. This indicates the high probability of deposition of these particles on the side walls. 

This trend of particle size effect, while particle density, flow and electric fields remain constant can be mapped as a straight line parallel to $d_p$ axis (marked in the transition between regime B\&C), in figure \ref{fig:ModesPraticleTraj}. Hence when such straight line crosses regime C (inertial deposition) of $St_v\gtrsim 1$ and $n_D\leq 0.1$, while transiting from regime B (ILC) to regime D (ES capture), particle size selective deposition arises for $\frac{18 \mu_a D}{\rho_p v_0} \lesssim d_p^2 \lesssim 0.1 \frac{12 \mu_a}{\epsilon_0 \epsilon_a \kappa_{CM}} \frac{D v_0}{E_0^2}$. According to this scaling, particle selection window for the parameters of case 4 of figure \ref{fig:SegregationChart}A are $12\mu m$ to $28\mu m$, which is of the order of the numerical results. Hence selective deposition arises when, $E_0^2 < 0.1 \frac{2 \rho_p}{3 \epsilon_0 \epsilon_a \kappa_{CM}} v_0^2$. Operating conditions can be manipulated as, $v_0 = \frac{18 \mu_a D}{\rho_p d_{p(l)}^2}$ and $E_0^2 = 0.1 \frac{12 \mu_a}{\epsilon_0 \epsilon_a \kappa_{CM}} \frac{D v_0}{d_{p(u)}^2}$to collect particles of size bound between $d_{p(l)}$ and $d_{p(u)}$. Fraction of particles deposited on the walls as a function of their size  is shown in fig \ref{fig:SegregationChart}C. This indicates manipulation of particle size selection through the operating parameters.
\subsection{\label{subsec:anal_charge}Dynamics of particles with mild charge}
\begin{figure}[!htbp]
\includegraphics[width=0.75\textwidth]{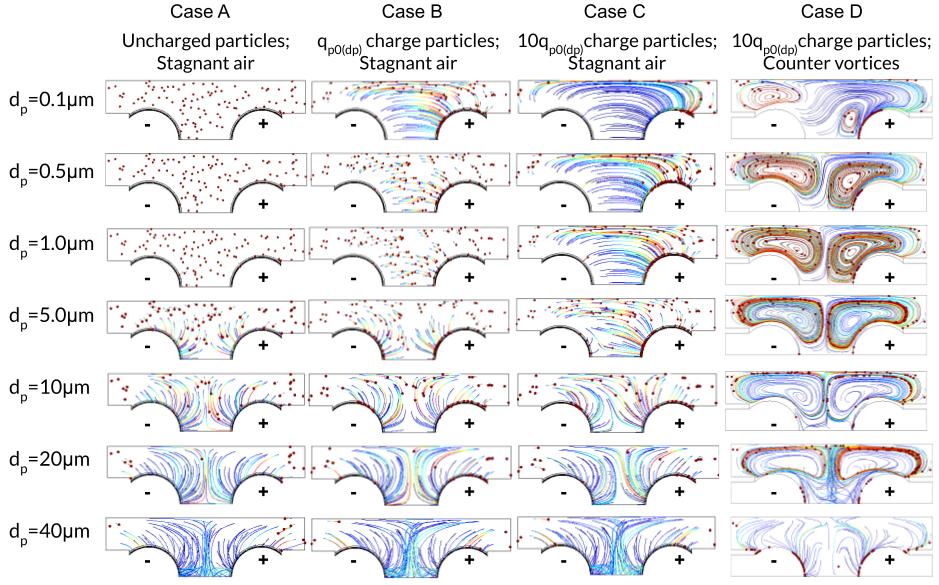}
\caption[Charged particle trajectories in flow and electric fields between electrodes]{\label{fig:chargedTraj}Trajectories of charged particle of density $1200kg/m^3$ between $0.5mm$ diameter electrodes of $400V$ potential difference, under different cases of particle charge and air circulations.}
\end{figure}
\begin{figure}[!htbp]
\includegraphics[width=0.5\textwidth]{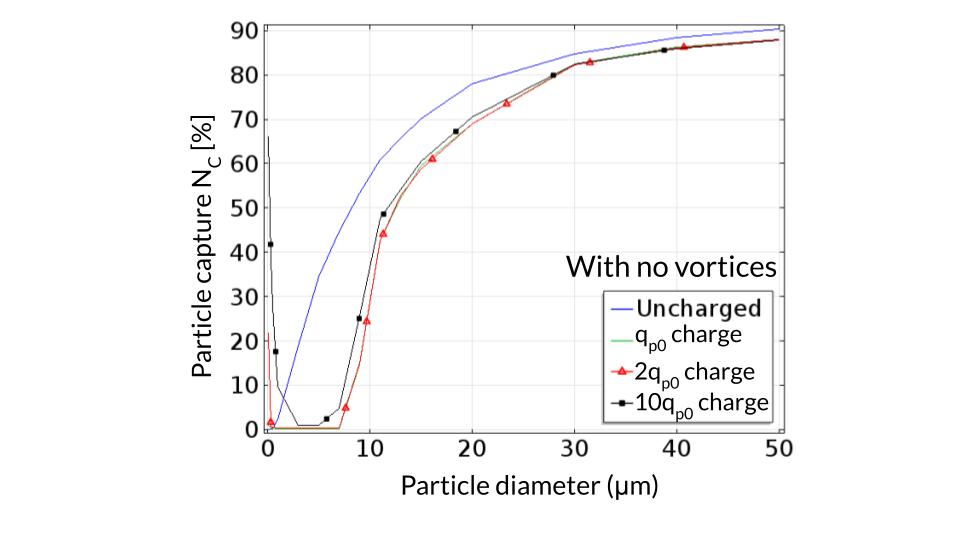}
\caption[Influence of charge on electrostatic particle capture without vortices]{\label{fig:chargeESP} Influence of particle size on electrostatic particle capture in stagnant air for particles of density $1200kg/m^3$with different magnitudes of their size dependant charge subject to electric field between electrodes of  0.5mm diameter at 400V potential difference.}
\end{figure}

Inherent charge on particles of $0.3\mu m$ and $10 \mu m$ diameter are estimated as 1 and 5 electrons respectively, inline with aerosol literature discussed in section \ref{sec:motivation}\cite{mayya1994charging,davenport1978field,kousaka1981measurement}. Charge on particle of any other size is approximated linearly interpolated and extrapolated as $q_{p0(dp)}=-\left(4 \left( \frac{d_p-0.3[\mu m]}{9.7\mu m}\right)+1\right)$. Figure \ref{fig:chargedTraj} lists trajectories of particles in different cases, with difference unipolar charge magnitudes in the electric field. Case A is the uncharged particles in stagnant air, where influence of DEP is symmetric in the domain and increases with the particle size. As the charge on the particles increases to $q_p= q_{p0(dp)}$ and $q_p= 10 q_{p0(dp)}$ for case B \& C capture due to Coulombic force appears for submicron particles. However the unipolar charge breaks the symmetry of particles trajectories hence slows down the capture for intermediate sizes, as shown in figure \ref{fig:chargeESP}, while further larger particles are dominated by the DEP capture. Charged particles in the electric field are influenced by counter vortices, in case D. Submicron particles in this case are strongly affected by both Coulombic force and aerodynamic drag. Their antagonizing influences near the co-polarity electrode results in unique closed airborne trajectories. On the other hand, larger particles in this case D behave more like uncharged particles.
\section{\label{sec:conclusions}Conclusions and remarks}
\begin{enumerate}
\item \textit{Co-flow capture enhancement}: When the air flow in front of the electrodes is inline with the direction of electrostatic particle attraction, capture efficiency of micron sized particles improves to be comparable with that of high particle charge in typical air ionization ion injection case in ESPs, as shown in the figure \ref{fig:counterVort_capture}.
\item \textit{Counter flow cut-off of particle capture}: A phase diagram with 6 states of particle behavior are identified in case of counter circulation vortices between electrodes, at different regimes of varying dominance of drag-inertia-DEP mapped on dimensionless numbers ($St_v, n_D$) as shown in figure \ref{fig:ModesPraticleTraj}. Cut-off particle capture is identified as a transition between drag-inertial regime-B to DEP dominant regime-D. Combined influence of electrostatic and air flow operating conditions are quantified as a single dimensionless number $\xi_v $, the ratio of electrostatic and inertial influences on the uncharged particles. At $\xi_v =\frac{n_D}{St_v}\sim 0.2$ vortical flow cutoff of particle capture abruptly shifts to DEP dominant capture of fine micron particles, hence might be wanted regime to avoid deterioration of particle capture in electrostatic precipitators. 
\item \textit{Segregation and selective collection of particles}: Particles in a polydisperse aerosol are separated spatially depending on their regime of location on the phase diagram of counter vortices subject to electric field, figure \ref{fig:ModesPraticleTraj}. Operating flow velocity and electric field magnitude parameters can be adjusted to selectively deposit particles bound with in a certain size and inertia window, on to the domain side walls, as demonstrated in figure \ref{fig:SegregationChart} C.
\item \textit{Particle flight manipulation}: Inertial limit cycle trapping of particles is identified in counter circulation vortices around curved surfaces, which can also segregates particles in their airborne trajectories. However such segregation may be suppressed by strong applied electric field.
\item \textit{Mild charge influence:} Unipolar charge on particles breaks symmetry of trajectories in the domain between the electrodes, which may slowdown their electrostatic capture.
\end{enumerate}

Hence, combining the electric and flow fields through suitable design of configuration can be used to improve the particulate capture, selective collection and active manipulation of airborne particulate flight. Experimental validation of these results in similar configuration of flow and electric fields, as suggested in the appendix \ref{append:vortexCreationMethods} and theoretical generalization to extend these results for wider configurations of orientation between the air flow and electric fields would be significant for further progress in this area of future applications. Efficiency of an ESP for particulate capture may be improved by utilizing co-circulation of air and systematically avoiding counter-circulations in the design of the system. Leveraging air circulations to replace plasma arc ion injection in to air would be of crucial role in improving ESP adaptation under health and environmental impacts. Selective deposition of particles identified at the interplay of counter circulating air flow and electrostatic fields might be combined with microfluidic biosampling devices\cite{hong2015continuous,schaap2012transport}, wet wall \cite{mcfarland2010wetted,lin2010efficient} and membrane wall\cite{phadke2021novel,bayless2004membrane} airborne particulate collectors for novel active, faster and reliable bioaerosol sampling techniques. Inertial limit cycle manipulation of airborne particles could be of importance in particulate flight through narrow irregular channels, such as aerosol drug delivery applications. Realizing such particle trapping in airborne state around electrodes could be used as aerosol tweezers. With such wide future prospects, much of the mechanics and opportunities at the intersection of aerodynamic and electrostatic influences on airborne particulate is yet to be explored.
\section{Author declarations }
\subsection{Conflicts of interest}
The authors have no conflicts of interest.
\subsection{Authors' contributions}
PS conceptualized and conducted research. RT supervised and reviewed to improve quality and completeness of the work.
\subsection{Acknowledgements}
Authors thank Dr.Y S Mayya for his guidance and suggestions on the work from aerosol engineering perspective. PS acknowledges Department of Science and Technology (DST), India for financial support during the period of research.
\subsection{Data availability}
The data that support the findings of this study are available from the corresponding author
upon reasonable request.
\newpage
\appendix
\section{\label{append:crescentModel} Particles in crescent model vortex}

\subsection{\label{append:subsec:model_particles_in_circular_vortex} Airborne particles in circular arc streamline flow}
A point mass particle is considered to be placed in a circular arc concentric streamlines of velocity $\textbf{v}\equiv (v_t,v_n)$ with uniform angular velocity $\omega$. Components of the velocity normal and tangential to the streamlines are,
\begin{align*}
    v_n &=0\\
    v_t &= r \omega
\end{align*}Particle position, velocity and angular velocity vectors with respect to the center of the vortex are expressed as $\textbf{r}_p \equiv (r_p,\theta_p)$, $\textbf{v}_p \equiv (v_{pt},v_{pn})$ and $\omega_p$. Then equation of particle motion in terms of time derivatives of position vector $\textbf{r}_p$ in noninertial rotating frame of reference of the circular vortex can be expressed as\cite{synge2011principles},
\begin{align}
 \textbf{v}_p= \pdv{\textbf{r}_p}{t} &= \pdv{{r_p}}{t} \hat{\textbf{r}} + \overline{\omega}_p x \textbf{r}_p 
 \end{align}velocity with radial and tangential components on the RHS and
 \begin{align}
\textbf{a}_p =\pdv[2]{\textbf{r}_p}{t} &= \pdv[2]{r_p}{t} \hat{\textbf{r}}+ \overline{\omega}_p x (\overline{\omega}_p x\textbf{r}_p) + \left( \pdv{\overline{\omega}_p}{t}x\textbf{r}_p \right) + 2 \overline{\omega}_p
x  \pdv{{r}_p}{t}\hat{\textbf{r}} 
\end{align}
Particle acceleration above has the following terms on RHS. Radial acceleration components due to change in radial position and centrifugal force due to change in instantaneous velocity direction; and tangential components of Euler force due to change angular velocity and Coriolis force due to radial velocity of particle, respectively.

\begin{figure}[!htbp]
\includegraphics[width=\textwidth]{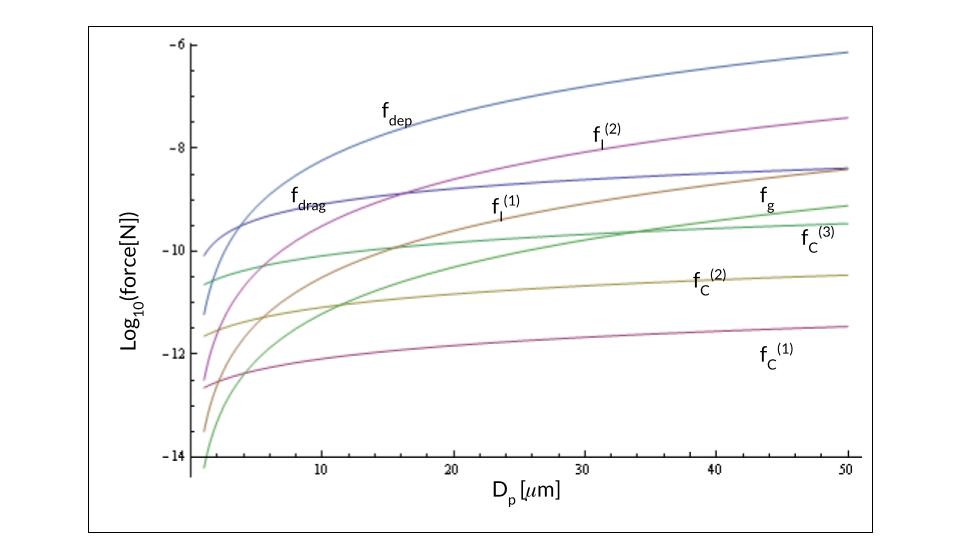}
\caption[Comparison of forces on aerosol particle]{\label{fig:DEP_dragForces} Variation of various forces, at the operating parameter scales, on the particle with its size $d_p$. $f_{dep}$ is dielectrophoretic force and $f_{drag}$ drag force. $f_C^{(1)},f_C^{(2)},f_C^{(3)}$ are Coulombic force with particle charge 1, 10 and 100 times the size dependent natural charge approximation $q_{p0}=-e \left(1+4 \left( \frac{d_p[\mu m]-0.3}{9.7}\right) \right)$ and $f_I^{(1)},f_I^{(2)}$ are inertial forcing response of particle of density $120kg/m^3, 1200kg/m^3$ respectively. $f_g$ is the gravitational force on particle of density $1200kg/m^3$. }
\end{figure}

Scaling this equation as discussed in section \ref{sec:method} and $\omega_p \sim \frac{1}{\tau_0}$,
\begin{align}
    St_v \pdv{\tilde{\textbf{v}}_p}{\tilde{t}} &= St_v  \pdv[2]{\tilde{r}_p}{\tilde{t}} \hat{\textbf{r}}+ St_v \overline{\tilde{\omega}}_px (\overline{\tilde{\omega}}_px\tilde{\textbf{r}}_p)+St_v \pdv{\overline{\tilde{\omega}}_{p}}{\tilde{t}}x \tilde{\textbf{r}}_p  + 2 St_v (\overline{\tilde{\omega}}_px  \pdv{\tilde{r}_p}{\tilde{\theta}_p} \tilde{\omega}_p) 
\end{align}Hence equation of motion of particle can be written as, 
\begin{align}
    St_v  \pdv[2]{\tilde{r}_p}{\tilde{t}} \hat{\textbf{r}} + St_v \overline{\tilde{\omega}}_px (\overline{\tilde{\omega}}_px \tilde{\textbf{r}}_p) +St_v \pdv{\overline{\tilde{\omega}}_{p}}{\tilde{t}}x \tilde{\textbf{r}}_p + 2 St_v (\overline{\tilde{\omega}}_px  \pdv{\tilde{r}_p}{\tilde{\theta}_p} \tilde{\omega}_p)   &= (\tilde{\textbf{v}}-\tilde{\textbf{v}}_p) + n_{D} \nabla \tilde{\textbf{E}}^2 + n_C \tilde{\textbf{E}}
\end{align} Variation of different forces, on the RHS, for particles of density $1200kg/m^3$, electrical permittivity of 10 and conductivity of $10^-4 S/m$ subject to typical values of an ESP, electric field of $E =10kV/cm$, electric field gradient of $\nabla E = 200kV/cm^2$ and air velocity of $1m/s$ is shown in figure \ref{fig:DEP_dragForces}. Gravitational influence is at least 1 order of magnitude smaller than vortical flow drag and 4 orders smaller than DEP force on particles.

Trajectories of particles in counter vortices subject to electric field in regime B of figure \ref{fig:ModesPraticleTraj} with mild inertial influence $St_v<1$ and weak electrostatic influence $n_D<<1$ and $n_C<<1$ is considered further. Since $St_v < 1$, centrifugal radial drift of particles in the concentric streamlines can be approximated to be gradual, i.e., $\pdv{\tilde{\textbf{r}}_p}{\tilde{\theta}_p}<<1$ hence Euler force is equal to the tangential acceleration force $St_v \pdv{\overline{\tilde{\omega}}_p}{t}x\tilde{\textbf{r}}_p \approx St_v \pdv{\tilde{v}_{pt}}{\tilde{t}}$. So the tangential direction momentum balance of particle is, 
\begin{align}
    St_v \pdv{\tilde{v}_{pt}}{\tilde{t}} &= (\tilde{v}_t-\tilde{v}_{pt})\\
\implies \tilde{v}_{pt} &= \tilde{v}_t (1-e^{-(\frac{2\tilde{t}}{St})})
\end{align}
For $St_v\leq0.1$, $\tilde{\textbf{v}}_{pt} \approx \tilde{\textbf{v}}_t$ hence $\overline{\omega}_{p}\approx\overline{\omega}$.
Hence scalar equation of particle motion in the normal direction to the circular streamlines is, 
\begin{align}
    St_v \pdv[2]{\tilde{r}_p}{\tilde{t}} = (\tilde{v}_n-\tilde{v}_{pn}) + St_v \tilde{r}_p \tilde{\omega}_p^2
\end{align}
\begin{align}
    St_v \pdv[2]{\tilde{r}_p}{\tilde{t}}+\pdv{\tilde{r}_p}{\tilde{t}} - St_v \tilde{r}_p \tilde{\omega}_p^2 &= 0
\end{align}
Hence, 
\begin{align}
    \tilde{r}_p &= \tilde{r}_{p(0)} e^{t/\tau_{pn}}
\end{align}
Where, $\tau_{pn}= \frac{2 St_v}{\sqrt{1+4 St_v^2 \tilde{\omega}_p^2}-1}$. Also, since $\tilde{\theta}_p = \tilde{\omega}_p t \approx \tilde{\omega} t$,
\begin{align}
    \tilde{r}_p &= \tilde{r}_{p(0)} e^{\tilde{\theta}_p/\tilde{\omega} \tau_{pn}} 
\end{align}
\subsection{\label{append:subsec:CrescentModel} Particle dynamics in a crescent streamlines flow}
\subsubsection{Structure of crescent streamlines} Now consider the crescent model of vortex described in section \ref{subsec:MethodCrescent}. Any position on segment 2 can be represented by $(r_p^{(2)},\theta_p^{(2)})$ with respect to its center of curvature $O_2$, as shown in the figure \ref{fig:crescent_model}. Consider a particle which is initially stationary at point A, in the segment 1 on the boundary with the segment 2, moves to point B and enters the segment 2. Further it moves to point C to enter back in to segment 1, as shown in the figure \ref{fig:crescent_model}. With variables $y_B$ and $y_C$ defined as,
\begin{align}
    y_{B,C}^{(1)} &= R_v - r_{p(B,C)}^{(1)}\\
    y_{B,C}^{(2)} &= r_{p(B,C)}^{(2)}-R_D\\
\end{align}Points B' \& B'' identified such that, line $\overline{O_2B'B}$ is a radial to segment 2 streamline through point B and angle $\angle{PB''B}=\frac{\pi}{2}$, as shown in the figure \ref{fig:crescent_model}. From the triangle $\Delta O_1BO_2$ angle $PBB'= (\pi-\theta_{B}^{(1)}+\theta_{B}^{(2)})$. Hence relation between $y_{B}^{(1)}$ and $y_{B}^{(2)}$ can be identified as, 
\begin{align}
y_B^{(2)}  &= \frac{y_B'}{cos(\gamma)} \\
\text{and,  }
y_B' &= \frac{y_B^{(1)}}{cos(\pi-(\theta_{B}^{(1)}-\theta_{B}^{(2)})+\gamma)}
\end{align}Where $\gamma$ is the angle between the secant $\overline{PB}$ and tangent of segment 2 at point B and $\theta_{B}^{(1)}=\frac{\theta_1}{2}, \theta_{B}^{(2)}=\frac{\theta_2}{2}$ and $y'_B=|BB''|$ as shown in figure \ref{fig:crescent_model}. Thus,
\begin{align}
y_B^{(2)} &= y_B^{(1)} Sec(\pi-(\theta_1-\theta_2)/2+\gamma) Sec(\gamma)
\end{align}
Similarly, coordinates of $C$ can be converted. Hence, the exchange between the coordinates of the points $B\&C$ in the coordinate systems of the two segments is,
\begin{align}
r_{p(B)}^{(2)} &= (R_V-r_B^{(1)}) Sec(\pi-(\theta_1-\theta_2)/2+\gamma) Sec(\gamma) + R_D\\
r_{p(C)}^{(1)} &= R_v-(r_c^{(2)}-R_D) Cos(\pi-(\theta_1-\theta_2)/2+\gamma) Cos(\gamma)
\end{align}
\subsubsection{Particle movement in crescent vortex}
Since,
\begin{align}
    r_{p(B)}^{(1)} &= r_{p(A)}^{(1)} e^{\frac{\theta_1}{\omega_1 \tau_{pn1}}}
\end{align}
Also, as $St_v \leq 1$, particle directions can be approximated to change immediately with air flow at the intersection of the segments of the vortices to stay on the streamline. After traversing on segment 2, position C distance from segment 2 center $r_{p(c)}'$
\begin{align}
    r_{p(c)}^{(2)} &= r_{p(B)}^{(2} e^{(\frac{\theta_2}{\omega_2 \tau_{pn2}})}
\end{align}
Hence its distance from center of segment 1
\begin{align*}
    r_{p(c)}^{(1)} =& r_{p(A)}^{(1)} \left(e^{\frac{\theta_1}{\omega_1 \tau_{pn1}}+ \frac{\theta_2}{\omega_2 \tau_{pn2}}} \right) + \\
   & \left[R_v\left(1-e^{\frac{\theta_2}{\omega_2 \tau_{pn2}}} \right) + R_D\left( (Cos(\pi-(\theta_1-\theta_2)/2+\gamma) Cos(\gamma))- e^{\frac{\theta_2}{\omega_2 \tau_{pn2}}} \right) \right]
\end{align*}

\begin{align}
\implies   r_{p(c)}^{(1)} &= \alpha r_{p(A)}^{(1)} +\beta
\end{align}
with, $\alpha=\left(e^{\frac{\theta_1}{\omega_1 \tau_{pn1}}+ \frac{\theta_2}{\omega_1 \tau_{pn2}}} \right)$ and \\ $\beta =R_v\left(1-e^{\frac{\theta_2}{\omega_2 \tau_{pn2}}} \right) + R_D\left( (Cos(\pi-(\theta_1-\theta_2)/2+\gamma) Cos(\gamma))- e^{\frac{\theta_2}{\omega_2 \tau_{pn2}}} \right)$. $\tau_{pn1,2} = \frac{2 St{v1,2}}{\sqrt{1+ St_{v1,2}^2}-1}$. $\omega_1$ and $\omega_2$ are considered as constant values in each of the segments of the forced vortex, as in case of circular vortex. $\chi$  defined as,
\[\chi= \frac{\omega_1}{\omega_2} \] and since the time scales $\tau_p$ and $\tau_{0}$ are same on both the segments  $St_{v1}=St_{v2}$. 
\[ \delta = \frac{\tau_{pn1}}{\tau_{pn2}} = \frac{\sqrt{St_{v}^2 \omega_2^2+1}-1}{\sqrt{St_{v}^2 \omega_1^2+1}-1} \]
When segment 2 is smaller than that of segment 1, with $\theta_1 \sim \frac{3\pi}{2}$, $\theta_2\sim \frac{\pi}{2}$ , angle $\gamma$ can be consider as a small value, to approximate the above relations as,
\begin{align}
 r_{p(c)}^{(1)} =&  r_{p(A)}^{(1)} \left(e^{\frac{\theta_1}{\omega_1 \tau_{pn1}}+ \frac{\theta_2}{\omega_2 \tau_{pn2}}} \right) \\
 &+\left[R_v\left(1-e^{\frac{\theta_2}{\omega_2 \tau_{pn2}}} \right) + R_D\left( \frac{1}{\sqrt{2}}- e^{\frac{\theta_2}{\omega_2 \tau_{pn2}}} \right) \right]
\end{align}
\subsubsection{Radial size of limit cycle trajectories}
\begin{figure}[!htbp]
\includegraphics[width=0.7\textwidth] {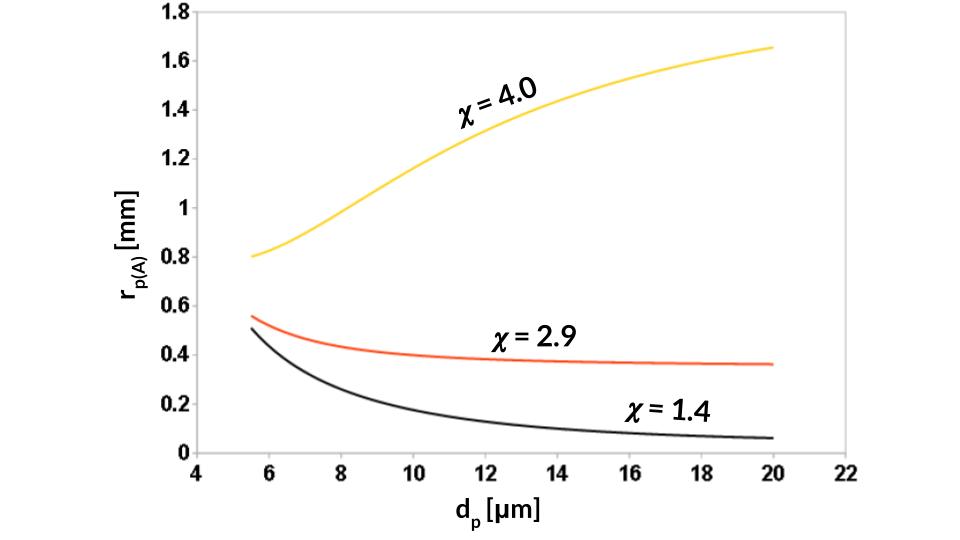}
\caption{\label{fig:crescent_Append}
Two trends of limit cycle trajectories size variation with particle size, for different values of  $\chi$, the parameter representing velocity distribution between segments of the crescent vortex.}
\end{figure}

Due to the opposite curvatures of the streamlines in the two segments, particle deviation from the streamlines is away from and towards the vortex center $O_1$ in the segment 1 \& 2 respectively. This results in the possibility for stable limit cycle trajectory of particle to exist, with the net radial displacement of the particle over a complete circulation being zero. Condition for such a Limit cycle of particle trajectories to exist is, 
\begin{align}
r_{p(A)}^{(1)}=
\begin{cases}
< r_{p(C)}^{(1)} \quad &; r_{p(A)}^{(1)}< r_{p(LC)}^{(1)}\\
= r_{p(C)}^{(1)} \quad &; r_{p(A)}^{(1)}= r_{p(LC)}^{(1)}\\
> r_{p(C)}^{(1)} \quad &; r_{p(A)}^{(1)}>r_{p(LC)}^{(1)}
\end{cases}
\end{align}
for $r_{p(LC)}>0$ is the value of $r_{p(A)}^{(1)}=r_{p(C)}^{(1)}$ on the limit cycle trajectory. Hence,
\begin{align}
    r_{p(LC)} &= \frac{\beta}{1-\alpha}
\end{align}
Solution to this equation indicates two possible trends of particle segregation in the crescent vortex for different values of $\chi$, as shown in the figure\ref{fig:crescent_Append}.
\begin{itemize}
    \item [Case 1] Limit cycle trajectory of particles of higher inertia is farther from the vortex center. Which is when the influence of segment 1 is dominates that of segment 2.
    \item [Case 2] Particles of higher inertia are pushed closer to the vortex center than the lighter particles. Which is when the influence of segment 2 is dominates that of segment 1.
\end{itemize}
Above Case 2 is observed for the parameter regime analyzed in this report figure \ref{fig:ILC}A, for $\theta_1=\frac{3 \pi}{2}$, $R_v=0.1 mm$ and $R_D=0.25 mm$, corresponding to $\chi=0.5$.
\newpage
\section{\label{append:vortexCreationMethods}Vortices generation around convex surfaces}
\begin{figure}[!htbp]
\includegraphics[width=0.7\textwidth] {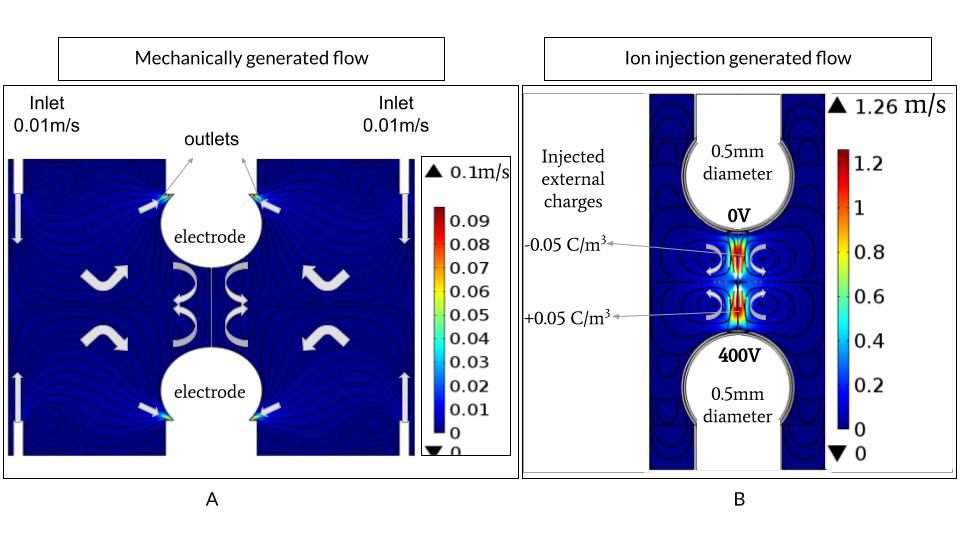}
\caption{\label{fig:vortexGeneration_Solid_Append}
Vortex generation between solid electrodes (A). Through mechanically generated external air flow (B). Controlled injection of externally generated ions between electrodes}
\end{figure}

\begin{figure}[!htbp]
\includegraphics[width=0.7\textwidth] {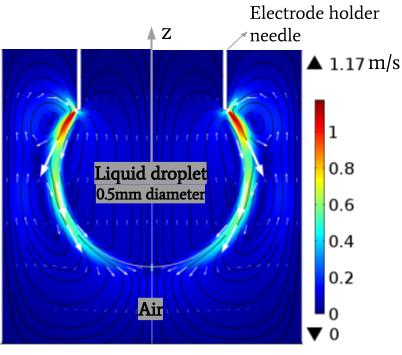}
\caption{\label{fig:MarangoniVortex_Append}
Vortex generation using surfactant driven Marangoni flow on a droplet}
\end{figure}

Five conceptual ideas to realize vortical flows and electrical fields of similar structure and strengths described in the section \ref{sec:problem}, shown in figure \ref{fig:geometry,vortices} are qualitatively discussed in this appendix, which might guide future work to validate the results experimentally.
First, mechanically externally generated air flow circulations with suitable inlet and outlet of air flow around the curved electrodes, as shown in figure \ref{fig:vortexGeneration_Solid_Append}A. Second, electrohydrodynamic gas flows created by injecting mild charges between the electrodes, as shown in the figure\ref{fig:vortexGeneration_Solid_Append}B. Air circulations of structure and magnitude similar to those in this work can be created with controlled injection of externally generated mild charges of about $\pm 0.05C/m^3$, which is about 1\% of the charge created by ionizing air in the whole domain between the electrodes. Due to Coulombic force exerted on the ions and thus on its surrounding air such air flows are created. Ionizing air guns are frequently used for generating and injecting ions into air.
Third, tangential force boundary condition used in the section \ref{sec:problem} to create circulations also closely resembles with interfacial Marangoni tangential flow created on droplets. Such flow on the droplet can lead to shear driven air flow creating circulations. Figure \ref{fig:MarangoniVortex_Append} indicates a droplet anchored on to a needle. Surfactant is dripped from the needle on to the droplet surface, creating interfacial tension(IFT) difference on the drop surface. Numerical simulation of the drop is conducted with the IFT varying from $72mN/m$ at the tip to $36mN/m$ at the anchored point, linearly with the axial length from the tip. Level set method is used to simulate the two phase system with such interfacial tension force, implemented on Comsol multiphysics finite element methods package. Resulting flow circulations in the air and droplet are as shown in the figure \ref{fig:MarangoniVortex_Append}. Electric potential is applied on the needle holding the droplet and a plate electrode opposite to it (on the bottom boundary of figure \ref{fig:MarangoniVortex_Append}).

Forth, \textit{Electrostatic fluid accelerator} are used in dielectric breakdown (DBD) cooling techniques on electronics and gas turbines to create strong local flows originated on solid surfaces to break the boundary layer on them, thus improve heat transfer between the solid surface and ambient gas flow \cite{benard2014electrical,moreau2007airflow}. Such flows are created by  dielectric breakdown between multiple sets of microscopic electrode pairs on the cooling solid surface. Similar surface originated tangential flows can be implemented on the electrode at a base voltage to create the required vortices and electric fields. Lastly, Srinivasula et.al.\cite{srinivasula2022EHDblower}, demonstrated strong local air circulations created by a pendant droplet electrohydrodynamic oscillations at low electric fields. Srinivasula et.al.\cite{srinivasula2022dropletairpurif}, adapted such technique to create circulations between array of large number of alternating rows of oscillating and stagnant liquid droplet electrodes to capture airborne particulate. Similar setup of solid spherical electrode pair between two oscillating liquid pairs can be configured to created circulations around the solid electrode surfaces.


\newpage
\bibliography{apssamp}

\begin{thebibliography}{10}

\bibitem{twomey1977atmospheric}
S.~Twomey, ``Atmospheric aerosols,'' {\em Aerosols}, 1977.

\bibitem{acheson1991elementary}
D.~J. Acheson, ``Elementary fluid dynamics, page. 175,'' 1991.

\bibitem{davies1953separation}
C.~Davies, ``The separation of airborne dust and particles,'' {\em Proceedings
  of the Institution of mechanical engineers}, vol.~167, no.~1b, pp.~185--213,
  1953.

\bibitem{rhodes2008introduction}
M.~J. Rhodes, {\em Introduction to particle technology}.
\newblock John Wiley \& Sons, 2008.

\bibitem{pich2017gas}
J.~Pich, ``Gas filtration theory,'' in {\em Filtration}, pp.~1--132, Routledge,
  2017.

\bibitem{friedlander1977smoke}
S.~K. Friedlander, ``Smoke, dust and haze: Fundamentals of aerosol behavior,''
  {\em New York}, 1977.

\bibitem{colgate1967enhanced}
S.~A. Colgate, ``Enhanced drop coalescence by electric fields in equilibrium
  with turbulence,'' {\em Journal of Geophysical Research}, vol.~72, no.~2,
  pp.~479--487, 1967.

\bibitem{gabyshev2020condensational}
D.~N. Gabyshev, A.~A. Fedorets, and O.~Klemm, ``Condensational growth of water
  droplets in an external electric field at different temperatures,'' {\em
  Aerosol Science and Technology}, vol.~54, no.~12, pp.~1556--1566, 2020.

\bibitem{yamamoto1981electrohydrodynamics}
T.~Yamamoto and H.~Velkoff, ``Electrohydrodynamics in an electrostatic
  precipitator,'' {\em Journal of fluid mechanics}, vol.~108, pp.~1--18, 1981.

\bibitem{lu2017analysis}
C.~Lu, C.~Yi, R.~Yi, and S.~Liu, ``Analysis of the operating parameters of a
  vortex electrostatic precipitator,'' {\em Plasma Science and Technology},
  vol.~19, no.~2, p.~025504, 2017.

\bibitem{jaworek2007modern}
A.~Jaworek, A.~Krupa, and T.~Czech, ``Modern electrostatic devices and methods
  for exhaust gas cleaning: A brief review,'' {\em Journal of electrostatics},
  vol.~65, no.~3, pp.~133--155, 2007.

\bibitem{parker1997electrostatic}
K.~R. Parker, ``Why an electrostatic precipitator?,'' in {\em Applied
  Electrostatic Precipitation}, pp.~1--10, Springer, 1997.

\bibitem{mizuno2000electrostatic}
A.~Mizuno, ``Electrostatic precipitation,'' {\em IEEE Transactions on
  Dielectrics and Electrical Insulation}, vol.~7, no.~5, pp.~615--624, 2000.

\bibitem{boelter1997ozone}
K.~J. Boelter and J.~H. Davidson, ``Ozone generation by indoor, electrostatic
  air cleaners,'' {\em Aerosol science and technology}, vol.~27, no.~6,
  pp.~689--708, 1997.

\bibitem{chen2002ozone}
J.~Chen and J.~H. Davidson, ``Ozone production in the positive dc corona
  discharge: Model and comparison to experiments,'' {\em Plasma chemistry and
  plasma processing}, vol.~22, no.~4, pp.~495--522, 2002.

\bibitem{tepper2007electrospray}
G.~Tepper, R.~Kessick, and D.~Pestov, ``An electrospray-based, ozone-free air
  purification technology,'' {\em Journal of Applied Physics}, vol.~102,
  no.~11, p.~113305, 2007.

\bibitem{mainelis2020bioaerosol}
G.~Mainelis, ``Bioaerosol sampling: Classical approaches, advances, and
  perspectives,'' {\em Aerosol Science and Technology}, vol.~54, no.~5,
  pp.~496--519, 2020.

\bibitem{andersen1958new}
A.~A. Andersen, ``New sampler for the collection, sizing, and enumeration of
  viable airborne particles,'' {\em Journal of Bacteriology}, vol.~76, no.~5,
  pp.~471--484, 1958.

\bibitem{nevalainen1992performance}
A.~Nevalainen, J.~Pastuszka, F.~Liebhaber, and K.~Willeke, ``Performance of
  bioaerosol samplers: collection characteristics and sampler design
  considerations,'' {\em Atmospheric Environment. Part A. General Topics},
  vol.~26, no.~4, pp.~531--540, 1992.

\bibitem{dolovich2011aerosol}
M.~B. Dolovich and R.~Dhand, ``Aerosol drug delivery: developments in device
  design and clinical use,'' {\em The Lancet}, vol.~377, no.~9770,
  pp.~1032--1045, 2011.

\bibitem{pleasants2018aerosol}
R.~A. Pleasants and D.~R. Hess, ``Aerosol delivery devices for obstructive lung
  diseases,'' {\em Respiratory care}, vol.~63, no.~6, pp.~708--733, 2018.

\bibitem{owen1992airborne}
M.~Owen, D.~Ensor, and L.~Sparks, ``Airborne particle sizes and sources found
  in indoor air,'' {\em Atmospheric Environment. Part A. General Topics},
  vol.~26, no.~12, pp.~2149--2162, 1992.

\bibitem{hinds1999aerosol}
W.~C. Hinds, {\em Aerosol technology: properties, behavior, and measurement of
  airborne particles, Chapter 3}.
\newblock John Wiley \& Sons, 1999.

\bibitem{morsi1972investigation}
S.~Morsi and A.~Alexander, ``An investigation of particle trajectories in
  two-phase flow systems,'' {\em Journal of Fluid mechanics}, vol.~55, no.~2,
  pp.~193--208, 1972.

\bibitem{mainelis2002effect}
G.~Mainelis, R.~L. G{\'o}rny, T.~Reponen, M.~Trunov, S.~A. Grinshpun, P.~Baron,
  J.~Yadav, and K.~Willeke, ``Effect of electrical charges and fields on injury
  and viability of airborne bacteria,'' {\em Biotechnology and bioengineering},
  vol.~79, no.~2, pp.~229--241, 2002.

\bibitem{liang1994characteristics}
W.-J. Liang and T.~Lin, ``The characteristics of ionic wind and its effect on
  electrostatic precipitators,'' {\em Aerosol Science and Technology}, vol.~20,
  no.~4, pp.~330--344, 1994.

\bibitem{podlinski2006electrohydrodynamic}
J.~Podli{\'n}ski, J.~Dekowski, J.~Mizeraczyk, D.~Brocilo, and J.-S. Chang,
  ``Electrohydrodynamic gas flow in a positive polarity wire-plate
  electrostatic precipitator and the related dust particle collection
  efficiency,'' {\em Journal of Electrostatics}, vol.~64, no.~3-4,
  pp.~259--262, 2006.

\bibitem{chun2007numerical}
Y.~N. Chun, J.-S. Chang, A.~A. Berezin, and J.~Mizeraczyk, ``Numerical modeling
  of near corona wire electrohydrodynamic flow in a wire-plate electrostatic
  precipitator,'' {\em IEEE Transactions on Dielectrics and Electrical
  Insulation}, vol.~14, no.~1, pp.~119--124, 2007.

\bibitem{atten1987electrohydrodynamic}
P.~Atten, F.~M. McCluskey, and A.~C. Lahjomri, ``The electrohydrodynamic origin
  of turbulence in electrostatic precipitators,'' {\em IEEE Transactions on
  Industry Applications}, vol.~IA-23, no.~4, pp.~705--711, 1987.

\bibitem{soldati1998turbulence}
A.~Soldati and S.~Banerjee, ``Turbulence modification by large-scale organized
  electrohydrodynamic flows,'' {\em Physics of Fluids}, vol.~10, no.~7,
  pp.~1742--1756, 1998.

\bibitem{zhu2019numerical}
Y.~Zhu, M.~Gao, M.~Chen, J.~Shi, and W.~Shangguan, ``Numerical simulation of
  capture process of fine particles in electrostatic precipitators under
  consideration of electrohydrodynamics flow,'' {\em Powder Technology},
  vol.~354, pp.~653--675, 2019.

\bibitem{gao2020effect}
W.~Gao, Y.~Wang, H.~Zhang, B.~Guo, C.~Zheng, J.~Guo, X.~Gao, and A.~Yu,
  ``Effect of the vortex formed by the electrohydrodynamic flow on the motion
  of particles in a needle-plate electrostatic precipitator,'' {\em Aerosol and
  Air Quality Research}, vol.~20, no.~12, pp.~2911--2924, 2020.

\bibitem{britigan2006quantification}
N.~Britigan, A.~Alshawa, and S.~A. Nizkorodov, ``Quantification of ozone levels
  in indoor environments generated by ionization and ozonolysis air
  purifiers,'' {\em Journal of the Air \& Waste Management Association},
  vol.~56, no.~5, pp.~601--610, 2006.

\bibitem{yanallah2009experimental}
K.~Yanallah, F.~Pontiga, A.~Fernandez-Rueda, and A.~Castellanos, ``Experimental
  investigation and numerical modelling of positive corona discharge: ozone
  generation,'' {\em Journal of Physics D: Applied Physics}, vol.~42, no.~6,
  p.~065202, 2009.

\bibitem{kim2010electrospray}
J.-H. Kim, H.-S. Lee, H.-H. Kim, and A.~Ogata, ``Electrospray with
  electrostatic precipitator enhances fine particles collection efficiency,''
  {\em Journal of Electrostatics}, vol.~68, no.~4, pp.~305--310, 2010.

\bibitem{arumugham2013two}
A.~K. Arumugham-Achari, J.~Grifoll, and J.~Rosell-Llompart, ``Two-way coupled
  numerical simulation of electrospray with induced gas flow,'' {\em Journal of
  aerosol science}, vol.~65, pp.~121--133, 2013.

\bibitem{haig2016bioaerosol}
C.~Haig, W.~Mackay, J.~Walker, and C.~Williams, ``Bioaerosol sampling: sampling
  mechanisms, bioefficiency and field studies,'' {\em Journal of Hospital
  Infection}, vol.~93, no.~3, pp.~242--255, 2016.

\bibitem{jing2013microfluidic}
W.~Jing, W.~Zhao, S.~Liu, L.~Li, C.-T. Tsai, X.~Fan, W.~Wu, J.~Li, X.~Yang, and
  G.~Sui, ``Microfluidic device for efficient airborne bacteria capture and
  enrichment,'' {\em Analytical chemistry}, vol.~85, no.~10, pp.~5255--5262,
  2013.

\bibitem{collins2017selective}
D.~J. Collins, B.~L. Khoo, Z.~Ma, A.~Winkler, R.~Weser, H.~Schmidt, J.~Han, and
  Y.~Ai, ``Selective particle and cell capture in a continuous flow using
  micro-vortex acoustic streaming,'' {\em Lab on a Chip}, vol.~17, no.~10,
  pp.~1769--1777, 2017.

\bibitem{khojah2017size}
R.~Khojah, R.~Stoutamore, and D.~Di~Carlo, ``Size-tunable microvortex capture
  of rare cells,'' {\em Lab on a Chip}, vol.~17, no.~15, pp.~2542--2549, 2017.

\bibitem{mainelis2002induction}
G.~Mainelis, K.~Willeke, P.~Baron, S.~A. Grinshpun, and T.~Reponen, ``Induction
  charging and electrostatic classification of micrometer-size particles for
  investigating the electrobiological properties of airborne microorganisms,''
  {\em Aerosol science and technology}, vol.~36, no.~4, pp.~479--491, 2002.

\bibitem{kim2007micromachined}
Y.-H. Kim, J.-Y. Maeng, D.~Park, I.-H. Jung, J.~Hwang, and Y.-J. Kim,
  ``Micromachined cascade virtual impactor with a flow rate distributor for
  wide range airborne particle classification,'' {\em Applied Physics Letters},
  vol.~91, no.~4, p.~043512, 2007.

\bibitem{kang2014real}
J.~S. Kang, K.~S. Lee, S.~S. Kim, G.-N. Bae, and J.~H. Jung, ``Real-time
  detection of an airborne microorganism using inertial impaction and
  mini-fluorescent microscopy,'' {\em Lab on a Chip}, vol.~14, no.~1,
  pp.~244--251, 2014.

\bibitem{kauppinen1989static}
E.~I. Kauppinen, A.~V. J{\"a}ppinen, R.~E. Hillamo, A.~H. Rantio-Lehtim{\"a}ki,
  and A.~S. Koivikko, ``A static particle size selective bioaerosol sampler for
  the ambient atmosphere,'' {\em Journal of aerosol science}, vol.~20, no.~7,
  pp.~829--838, 1989.

\bibitem{hong2015continuous}
S.~C. Hong, J.~S. Kang, J.~E. Lee, S.~S. Kim, and J.~H. Jung, ``Continuous
  aerosol size separator using inertial microfluidics and its application to
  airborne bacteria and viruses,'' {\em Lab on a Chip}, vol.~15, no.~8,
  pp.~1889--1897, 2015.

\bibitem{schaap2012transport}
A.~Schaap, W.~C. Chu, and B.~Stoeber, ``Transport of airborne particles in
  straight and curved microchannels,'' {\em Physics of Fluids}, vol.~24, no.~8,
  p.~083301, 2012.

\bibitem{moon2009dielectrophoretic}
H.-S. Moon, Y.-W. Nam, J.~C. Park, and H.-I. Jung, ``Dielectrophoretic
  separation of airborne microbes and dust particles using a microfluidic
  channel for real-time bioaerosol monitoring,'' {\em Environmental science \&
  technology}, vol.~43, no.~15, pp.~5857--5863, 2009.

\bibitem{chan2006dry}
H.-K. Chan, ``Dry powder aerosol drug delivery—opportunities for colloid and
  surface scientists,'' {\em Colloids and Surfaces A: Physicochemical and
  Engineering Aspects}, vol.~284, pp.~50--55, 2006.

\bibitem{gharse2016large}
S.~Gharse and J.~Fiegel, ``Large porous hollow particles: lightweight champions
  of pulmonary drug delivery,'' {\em Current pharmaceutical design}, vol.~22,
  no.~17, pp.~2463--2469, 2016.

\bibitem{babincova2009magnetic}
M.~Babincova, P.~Babinec, {\em et~al.}, ``Magnetic drug delivery and targeting:
  principles and applications,'' {\em Biomed Pap Med Fac Univ Palacky Olomouc
  Czech Repub}, vol.~153, no.~4, pp.~243--50, 2009.

\bibitem{pourmehran2015simulation}
O.~Pourmehran, M.~Rahimi-Gorji, M.~Gorji-Bandpy, and T.~Gorji, ``Simulation of
  magnetic drug targeting through tracheobronchial airways in the presence of
  an external non-uniform magnetic field using lagrangian magnetic particle
  tracking,'' {\em Journal of Magnetism and Magnetic Materials}, vol.~393,
  pp.~380--393, 2015.

\bibitem{russel1991colloidal}
W.~B. Russel, W.~Russel, D.~A. Saville, and W.~R. Schowalter, {\em Colloidal
  dispersions}.
\newblock Cambridge university press, 1991.

\bibitem{norbury1973family}
J.~Norbury, ``A family of steady vortex rings,'' {\em Journal of Fluid
  Mechanics}, vol.~57, no.~3, pp.~417--431, 1973.

\bibitem{scase2018hill}
M.~Scase and H.~Terry, ``Hill's spherical vortex in a rotating fluid,'' {\em
  arXiv preprint arXiv:1801.09954}, 2018.

\bibitem{chapman1967formation}
D.~S. Chapman and P.~Critchlow, ``Formation of vortex rings from falling
  drops,'' {\em Journal of Fluid Mechanics}, vol.~29, no.~1, pp.~177--185,
  1967.

\bibitem{chen2003transient}
W.-H. Chen and Y.-C. Chung, ``Transient mass transfer and vortex bifurcation of
  an aerosol droplet in motion,'' {\em Aerosol Science \& Technology}, vol.~37,
  no.~8, pp.~640--658, 2003.

\bibitem{batchelor2000introduction}
C.~K. Batchelor and G.~Batchelor, {\em An introduction to fluid dynamics}.
\newblock Cambridge university press, page 526, 2000.

\bibitem{khillare2012airborne}
P.~S. Khillare and S.~Sarkar, ``Airborne inhalable metals in residential areas
  of delhi, india: distribution, source apportionment and health risks,'' {\em
  Atmospheric pollution research}, vol.~3, no.~1, pp.~46--54, 2012.

\bibitem{chong2013inertial}
K.~Chong, S.~D. Kelly, S.~Smith, and J.~D. Eldredge, ``Inertial particle
  trapping in viscous streaming,'' {\em Physics of Fluids}, vol.~25, no.~3,
  p.~033602, 2013.

\bibitem{chong2016transport}
K.~Chong, S.~D. Kelly, S.~T. Smith, and J.~D. Eldredge, ``Transport of inertial
  particles by viscous streaming in arrays of oscillating probes,'' {\em
  Physical Review E}, vol.~93, no.~1, p.~013109, 2016.

\bibitem{mayya1994charging}
Y.~Mayya, ``Charging-induced drift and diffusion of aerosol particles in
  oscillating electric fields,'' {\em Journal of aerosol science}, vol.~25,
  no.~2, pp.~277--288, 1994.

\bibitem{davenport1978field}
H.~M. Davenport and L.~K. Peters, ``Field studies of atmospheric particulate
  concentration changes during precipitation,'' {\em Atmospheric Environment
  (1967)}, vol.~12, no.~5, pp.~997--1008, 1978.

\bibitem{kousaka1981measurement}
Y.~KOUSAKA, K.~OKUYAMA, M.~ADACHI, and K.~EBIE, ``Measurement of electric
  charge of aerosol particles generated by various methods,'' {\em Journal of
  Chemical Engineering of Japan}, vol.~14, no.~1, pp.~54--58, 1981.

\bibitem{mcfarland2010wetted}
A.~R. McFarland, J.~S. Haglund, M.~D. King, S.~Hu, M.~S. Phull, B.~W. Moncla,
  and Y.~Seo, ``Wetted wall cyclones for bioaerosol sampling,'' {\em Aerosol
  Science and Technology}, vol.~44, no.~4, pp.~241--252, 2010.

\bibitem{lin2010efficient}
G.-Y. Lin, C.-J. Tsai, S.-C. Chen, T.-M. Chen, and S.-N. Li, ``An efficient
  single-stage wet electrostatic precipitator for fine and nanosized particle
  control,'' {\em Aerosol Science and Technology}, vol.~44, no.~1, pp.~38--45,
  2010.

\bibitem{phadke2021novel}
K.~S. Phadke, D.~G. Madival, J.~Venkataraman, D.~Kundu, K.~Ramanujan, N.~Holla,
  J.~Arakeri, G.~Tomar, S.~Datta, and A.~Ghatak, ``Novel non intrusive
  continuous use zebox technology to trap and kill airborne microbes,'' {\em
  Scientific reports}, vol.~11, no.~1, pp.~1--9, 2021.

\bibitem{bayless2004membrane}
D.~J. Bayless, M.~K. Alam, R.~Radcliff, and J.~Caine, ``Membrane-based wet
  electrostatic precipitation,'' {\em Fuel processing technology}, vol.~85,
  no.~6-7, pp.~781--798, 2004.

\bibitem{synge2011principles}
J.~L. Synge, {\em Principles of mechanics, Sec.12.3}.
\newblock Read Books Ltd, 2011.

\bibitem{benard2014electrical}
N.~Benard and E.~Moreau, ``Electrical and mechanical characteristics of surface
  ac dielectric barrier discharge plasma actuators applied to airflow
  control,'' {\em Experiments in Fluids}, vol.~55, no.~11, pp.~1--43, 2014.

\bibitem{moreau2007airflow}
E.~Moreau, ``Airflow control by non-thermal plasma actuators,'' {\em Journal of
  physics D: applied physics}, vol.~40, no.~3, p.~605, 2007.

\bibitem{srinivasula2022EHDblower}
P.~Srinivasula and D.~Biswal, ``Demonstration of droplet electrohydrodynamic
  blower in aerosol,'' {\em Manuscript submitted for publication}, 2022.

\bibitem{srinivasula2022dropletairpurif}
P.~Srinivasula and R.~Thaokar, ``A droplet based self cleaning electrostatic
  air cleaner,'' {\em Patent application submitted, Controller General of
  Patents, Designs and TradeMarks, India}, vol.~202121034284, 2022.

\end{thebibliography}

\end{document}